\documentclass{aa}  

\usepackage{graphicx}
\usepackage{txfonts}
\usepackage{hyperref}
\usepackage{multirow}
\usepackage{booktabs}
\usepackage{siunitx, array}
\usepackage{amsmath}
\usepackage{xcolor}
\usepackage{verbatim}
\usepackage{verbdef}
\usepackage{fancyvrb}
\newbox\verbbox

\begin{document}

\title{Impact of discontinuous grain size distributions\\on the spectral energy distribution of debris disks}
\titlerunning {Impact of grain size distributions on the SED of debris disks}
\author{
    M. Kim\inst{\ref{Warwick}, \ref{Warwick2}}
    \and
    S. Wolf\inst{\ref{Kiel}}}

\institute{
    \label{Warwick}Department of Physics, University of Warwick, Gibbet Hill Road, Coventry CV4 7AL, UK \\
    \email{{minjae.k.kim@warwick.ac.uk}}\and
    \label{Warwick2}Centre for Exoplanets and Habitability, University of Warwick, Gibbet Hill Road, Coventry CV4 7AL, UK
    \and
    \label{Kiel}Institute of Theoretical Physics and Astrophysics, University of Kiel, Leibnizstra\ss e 15, 24118 Kiel, Germany}

\date{Received 13 June, 2023; accepted XXXX XX, XXXX}

\abstract
{The collisional evolution of debris disks is expected to result in a  characteristic wavy pattern of the grain size distributions, i.e., an under/overabundance of particles of specific sizes. This perturbed grain size distribution potentially leaves characteristic patterns in the spectral energy distribution (SED) of the disk system.}
{We aim to quantify and understand the specific influence of discontinuous particle size distributions on the appearance of debris disks. For this purpose, we consider dust emission models based on two different grain size distributions, i.e., once with a single and once with a broken power law. In particular, the impact of an overabundance of small grains and an underabundance of medium-sized grains on the far-infrared/(sub-)millimeter regime on the dust reemission radiation and the potential to constrain discontinuities in the grain size distribution from (sub-)millimeter photometric measurements of debris disks are in the focus of our study.}
{We compare the spectral index $\alpha$ ($F_{\nu}\,\propto \nu^{\rm{\,\alpha}}$) in the case of a continuous grain size distribution with that of a discontinuous grain size distribution. We perform this comparison for central stars with different spectral types and two different disk structures (e.g., slim and broad debris dust rings).}
{Within the considered parameter space, we find a characteristic difference between the spectral slopes of the SED in the different scenarios. In particular, the overabundance of small grains resulting from collisional events and thus parameters defining the outcome of disk events in debris disks is potentially observable by comparison with the SED corresponding to a grain size distribution resulting from an ideal collisional cascade. More specifically, the overabundance of small grains leads to a steeper slope in the far-infrared/submillimeter regime, while the spectral index in the mm regime is hardly affected. On the other hand, the underabundance of medium-sized grains results in a slight steepening of the far-infrared slope of SED, while its primary effect is on the mm slope of SED, causing it to become shallower. We also find that the impact of an overabundance of small dust particles is more pronounced than that of an underabundance of medium-sized dust particles. Furthermore, we find that the difference between the spectral indices for the two different grain size distributions is largest for debris disks around brighter central stars and broader disks. However, the impact of the considered spatial distributions described by the fractional width of the disk system is weak. Our results also show that the dust composition is not the sole physical mechanism responsible for the spectral inversion observed in the far-IR to mm part of the SED of debris disk systems. Furthermore, the location of the spectral break is placed at different wavelength regimes if the grain size distribution is considered as a function of blowout size and stellar type.}
{}



\maketitle

\section{Introduction}\label{introduction}

The lifetime of small dust grains in debris disks is much shorter than the stellar age (\citealp{Artymowicz_Clampin1997}), indicating that dust cannot be primordial in these systems. Thus, the dust in debris disks must be continuously replenished through disruptive collisions. This process leads to the creation of fragments with a broad size distribution, which can be approximately described by a power law. For example, for a steady-state collisional cascade (\citealp{Dohnanyi1969}), the power law with $\gamma$ = -3.5 is valid for the differential size distribution, namely d$n(a)$ $\propto$ $a^\gamma$ d$a$, where $a$ is the size of the dust grain. However, due to the finite size distribution and size-dependent collisional strength of the colliding bodies/particles, the real slope of the size distribution may strongly deviate from this value and thus may not accurately reflect the true grain size distribution. For example, the collisional cascade does not extend down to arbitrarily small particles because tiny particles with a high $\beta$ ratio between radiation pressure and stellar gravity are quickly blown away by the stellar radiation pressure from the disks (\citealp{Krivov06}). Furthermore, nongravitational forces (e.g., Poynting-Robertson drag; \citealp{Backman1993}) acting on grains of particular sizes in the range of tens to hundreds of micrometers may further modify the grain size distribution (\citealp{Loehne2017}). 

In previous studies based on numerical simulation, a characteristic wavy pattern in the grain size distribution resulting from collisional evolution within debris disks was found (e.g., \citealp{Thebault_Augereau_2007, Kim2018}). These patterns, not solely described by a simple power law, but, e.g., by broken power laws, indicate the overabundance or underabundance of particles of specific sizes. In particular, the wavy pattern of the discontinuous grain size distribution becomes more pronounced (deeper) and broader in extreme scenarios involving lower material strength of dust grains (\citealp{Kim2018}). 

Furthermore, several observational findings have pointed toward the need for a more complex model for grain size distribution beyond a simple single power law. For instance, observational findings based on the comparison between flux and polarization measurements of protoplanetary disks at sub-millimeter (sub-mm) wavelengths have also raised questions regarding the corresponding grain size distribution (e.g., \citealp{Brunngraeber2021}). \citet{Buerger2023} also recently demonstrated an apparent overabundance of grains with radii between 3 and 6 mm in the derived grain size distribution of comet 67P/Churyumov-Gerasimenko. Additionally, \citet{Rodigas2015} found that the grain size distributions derived from the observation of thermal reemission radiation are not consistent with those from scattered light colors and/or scattering functions. Finally, \citet{Olofsson2023} recently demonstrated that the overabundance of small grains in debris disks could also explain an intriguing arc-like structure observed in total intensity. These findings suggest the presence of a discontinuous or more complicated grain size distribution, challenging the traditional single-power law model.

The grain size distribution directly influences the slope of the spectral energy distribution (SED) in the far-infrared (far-IR) to submillimeter/millimeter (mm) range of the thermal reemission by the dust (\citealp{Wyatt_Dent_2002}). Consequently, the presence of overabundant or underabundant particle sizes at these wavelengths potentially leaves an observable imprint in the SED. For example, the recent NIKA2 observation of HD 107146 indicates a break at the far-IR and sub-mm/mm wavelength regimes (\citealp{Lestrade2020}), which is further supported by supplementary observations, showing possibly a change in the slope of the SED (e.g., \citealp{Holland2017, Ricci2015b}). In this particular case, the spatial distribution of dust in this debris disk system cannot be a reason for this break at long wavelength because HD 107146 is observed as a narrow belt system (\citealp{Marino2018}). Moreover, the influence of the dust temperature on the Planck function is weak in the Rayleigh-Jeans regime, diminishing its importance at long wavelengths. However, given the small range of dust temperatures in a narrow belt system, the impact of the underlying grain size distribution becomes more pronounced. Consequently, a change in the SED slope due to an overabundance/underabundance of grains is expected in the wavelength region where grains of the corresponding size emit most efficiently.\newline

\noindent The goal of the current study is to evaluate the impact of the discontinuous size distribution of dust in debris disks on the SED of these systems. In turn, the results of this study are expected to serve as a tool to also address the opposite question, namely whether constraints on the parameters describing the collisions in debris disks can be derived from the analysis of the slope of their thermal reemission SED. The present study is structured as follows: In Section \ref{Modeling of debris disks}, we describe our model for the simulations of debris disk observables. In Section \ref{Results}, we quantify and discuss the impact of the grain size distributions on the observational appearance of debris disks, in particular, their SEDs (e.g., the slope of SEDs). In Section \ref{Discussion}, we discuss an application of our results to observations and possible degeneracies and connections to chemical and physical properties. We summarize our findings in Section \ref{Summary and Conclusions}.


\section{Modeling of debris disks}\label{Modeling of debris disks}

\subsection{Model parameters of debris disk system}\label{Debris disk model parameter}

The parameters and characteristics of stars, disks, and dust in our models are based on observations of currently known debris disk systems. Table~\ref{Disk and dust table} and \ref{Star table} summarize the detailed characteristics of disk and dust parameters.\newline

\begin{table*}[]
\centering
\def\arraystretch{1.3}                          
\caption{Disk and dust model parameters for the simulations and corresponding references.}
\label{Disk and dust table}                             
\begin{tabular}{ll}                           
\hline\hline                               
Parameter 								&         		Value \\
\hline
Narrow belt’s fractional width  & $\Delta$$R$/$R$ $\sim$ 0.17 \tablefootmark{\rm{\,a}}(\citealp{Lestrade_Thilliez_2015})\\ 
Broad belt’s fractional width & $\Delta$$R$/$R$ $\sim$ 1.20 \tablefootmark{\rm{\,b}} (\citealp{Marino_2020}) \\ \hline                                       
{Distance $d$}   &   7.7 pc (\citealp{Mamajek2012})\\
{Radial density distribution $n\,(r)$}  &	$n\,(r)$ $\propto$ $r^{\rm - 1.5}$ (\citealp{Krivov06})\\
\hline
{Dust composition \& bulk density} & Astrosil \& 3.5 g cm$^{-3}$ (\citealp{Draine2003})\\ 
{Minimum grain radius $a_{\rm{\,min}}$}  &  $a_{\rm\,bo}$ \tablefootmark{\rm{\,c}}(\citealp{Kirchschlager2013}) or 0.264~$\mu\rm{m}$ (\citealp{Kim2018}) \\
{Maximum grain radius $a_{\rm{\,max}}$}  &  10~cm (\citealp{Lestrade2020})\\
\hline                                          
\end{tabular}
\tablefoot{\tablefoottext{\rm{a, b, c}}{See Table~\ref{Star table}}}
\end{table*}

\begin{table*}[]
\centering
\def\arraystretch{1.3}                          
\caption{Stellar, disk, and dust model parameters for the simulations and their reference. The chosen stellar parameters are selected as one of the representative stars for each spectral type.}
\renewcommand{\arraystretch}{1.3}
\begin{tabular}{llllll}
\hline\hline
SpT                     		& Temperature                   	& Radius  & Narrow belt & Broad belt  &  Blow-out grain size \\
& [K] & [R$_\odot$] & $R_{\rm\, in}$, $R_{\rm\, out}$ [au] & $R_{\rm\, in}$, $R_{\rm\, out}$ [au] &  $a_{\rm\,bo}$ [$\mu$m]\\
\hline
A star ($\beta$ Pictoris)  		& 8052\tablefootmark{\rm{\,a}}    	& 1.8\tablefootmark{\rm{\,b}} & 92, 109    & 37, 163   & 1.99 \\
F star (HD 139664) 	& 6681\tablefootmark{\rm{\,c}}    	& 1.26\tablefootmark{\rm{\,d}} & 76, 90    & 31 , 136   & 1.05 \\
G star (HD 107146) 	& 5820\tablefootmark{\rm{\,e}}    	& 0.993\tablefootmark{\rm{\,e}} & 62, 73    & 25, 110  & 0.43 \\
K star (HD 53143)  	& 5224\tablefootmark{\rm{\,f}}    	& 0.85\tablefootmark{\rm{\,g}} & 57, 67    & 23, 101  & - \\
M star (AU mic)    		& 3500\tablefootmark{\rm{\,h}}    	& 0.5\tablefootmark{\rm{\,h}} & 38, 45    & 15, 68   & - \\
\hline
\end{tabular}
\tablefoot{\tablefoottext{\rm{a}}{\citealp{Gray2006}} \tablefoottext{\rm{b}}{\citealp{Kervella2004}} \tablefoottext{\rm{c}}{\citealp{Reiners2006}} \tablefoottext{\rm{d}}{\citealp{Rhee2007}} \tablefoottext{\rm{e}}{\citealp{Marino2018}} \tablefoottext{\rm{f}}{\citealp{Kalas2006}} \tablefoottext{\rm{g}}{\citealp{Watson11}} \tablefoottext{\rm{h}}{\citealp{Donati2023}}}
\label{Star table}
\end{table*}

\noindent\textbf{Disk Geometry}: We consider the inner and outer radii of disks based on the empirical relation between planetesimal belt radius and stellar luminosity (\citealp{Marta2018}) with two different belt fractional widths: a narrower belt (e.g., $\Delta$$R$/$R$ $\sim$ 0.17 of debris disks around $\varepsilon$ Eridani; \citealp{Lestrade_Thilliez_2015}) comparable to the classical Kuiper belt, and a broader belt (e.g., $\Delta$$R$/$R$ $\sim$ 1.25 of debris disks around HD 206893; \citealp{Marino_2020}). We note that broader belt systems show a broad range of dust temperatures. The mass of inner belts in a two-belt system is typically only a few percent compared to outer belts (e.g., \citealp{Ertel2011, Marino2018}), implying that the contribution of the inner belt to the slope of the SED is negligible. Thus, we consider a single belt system to reduce the complexity of the model. \newline

\noindent\textbf{Minimum/maximum grain size and dust composition}: For the smallest grain size, we apply a blow-out grain size $a_{\rm\,bo}$ (\citealp{Backman1993}) depending on the spectral type. For cases without a blowout size (i.e., for central stars' effective temperatures below 5250 K in our model setup; \citealp{Kirchschlager2013}), we consider a minimum grain size of 0.264 $\mu$m that has been used in previous studies (e.g., \citealp{Loehne2017, Kim2018}), which shows the similarity to the previous findings of debris disks around M-type stars (e.g., \citealp{Matthews2015}) based on the strong stellar wind (\citealp{Plavchan2005}), causing an effect similar to that of the radiation pressure force (\citealp{Augereau2006}). Furthermore, using Mie theory, silicate or ice grains comparable to this size are hardly affected by the stellar radiation pressure in debris disks around a solar-type star (\citealp{Cataldi2016}). We adopt a maximum grain radius of 10 cm (\citealp{Lestrade2020, Pawellek2021}) to analyze the dust continuum up to a wavelength of a few mm such as ATCA Survey (\citealp{Ricci2015b}). We note that the contribution of larger grains to the net flux is negligible when considering a steep grain size distribution. For the grain composition, we assume the simplest dust grains to be compact and homogenous spheres composed of astronomical silicate (referred to as "Astrosil") with a bulk density of 3.5 g cm$^{-3}$ and optical constants taken from \citet{Draine2003}. \newline

\noindent\textbf{Disk mass}: Based on previous surveys at sub-mm wavelengths, we consider a dust mass in debris disks of $\sim$ 10$^{\rm-7}$~${\rm M}_{\,\odot}$ for the single power law grain size distribution model (e.g., dust mass of most debris disks typically ranges from $\sim$ 10$^{\rm-9}$ to several 10$^{\rm-7}$\,${\rm M}_{\,\odot}$; \citealp{Greaves2005}). In the case of the broken power law grain size distribution model, the dust mass is varied depending on the scale factor but remains within a similar range (i.e., 1.0 to 1.26 $\times$ 10$^{\rm-7}$~${\rm M}_{\,\odot}$; see Table~\ref{Table:grain size distributions}). This is because the contribution of small dust grains mainly affects the dust reemission rather than significantly influencing the overall mass budget, which is dominated by larger grains. We consider the same mass of dust $M_{\rm{dust}}$ for both types of disk structures (i.e., narrow and broad belt-like disks). We note that within this model the overall mass of a disk does not significantly affect spectral index as long as the disk remains optically thin; however, a specific range of dust masses with the under/over-abundance of medium and small grains is indeed an important factor. The detailed characteristics of the modeled classical single power law and wavy pattern (i.e., broken power-law) grain size distribution are compiled in Fig.~\ref{fig:GDs} and Table~\ref{Table:grain size distributions}.\newline

\noindent\textbf{Central star}: We consider central stars of the following five spectral types (in brackets exemplary systems are listed): A star ($\beta$ Pictoris), F star (HD 139664), G star (HD 107146), K star (HD 53143), and M star (AU mic). Table~\ref{Star table} summarizes the detailed characteristics of stellar parameters and corresponding disk and dust parameters. \newline

\noindent\textbf{Simulation of SEDs}: For the calculation of model SEDs, we use the software tool {\it DMS} (\citealp{Kim2018}), which is optimized for optically thin emission of debris disks. Based on wavelength-dependent refractive indices, we calculate required the optical properties with the tool \textit{miex} on the basis of Mie scattering theory (\citealp{Wolf04}).

\subsection{Modeling of grain size distributions}\label{Gran size distribution}

To understand how the grain size distribution affects the SED of debris disks, we consider dust emission models that are based on two different grain size distributions: first a single power law grain size distribution (Sect.~\ref{Single power law GD}) and second a broken power law grain size distribution (Sect.~\ref{Broken power law GD}). Based on these distributions, we study the shape of the thermal reemission SED in the far-IR to sub-mm/mm wavelength range.

\subsubsection{Single power law grain size distribution}\label{Single power law GD}

\begin{figure}
\centerline\centering\includegraphics[width=9cm, height=5.7cm]{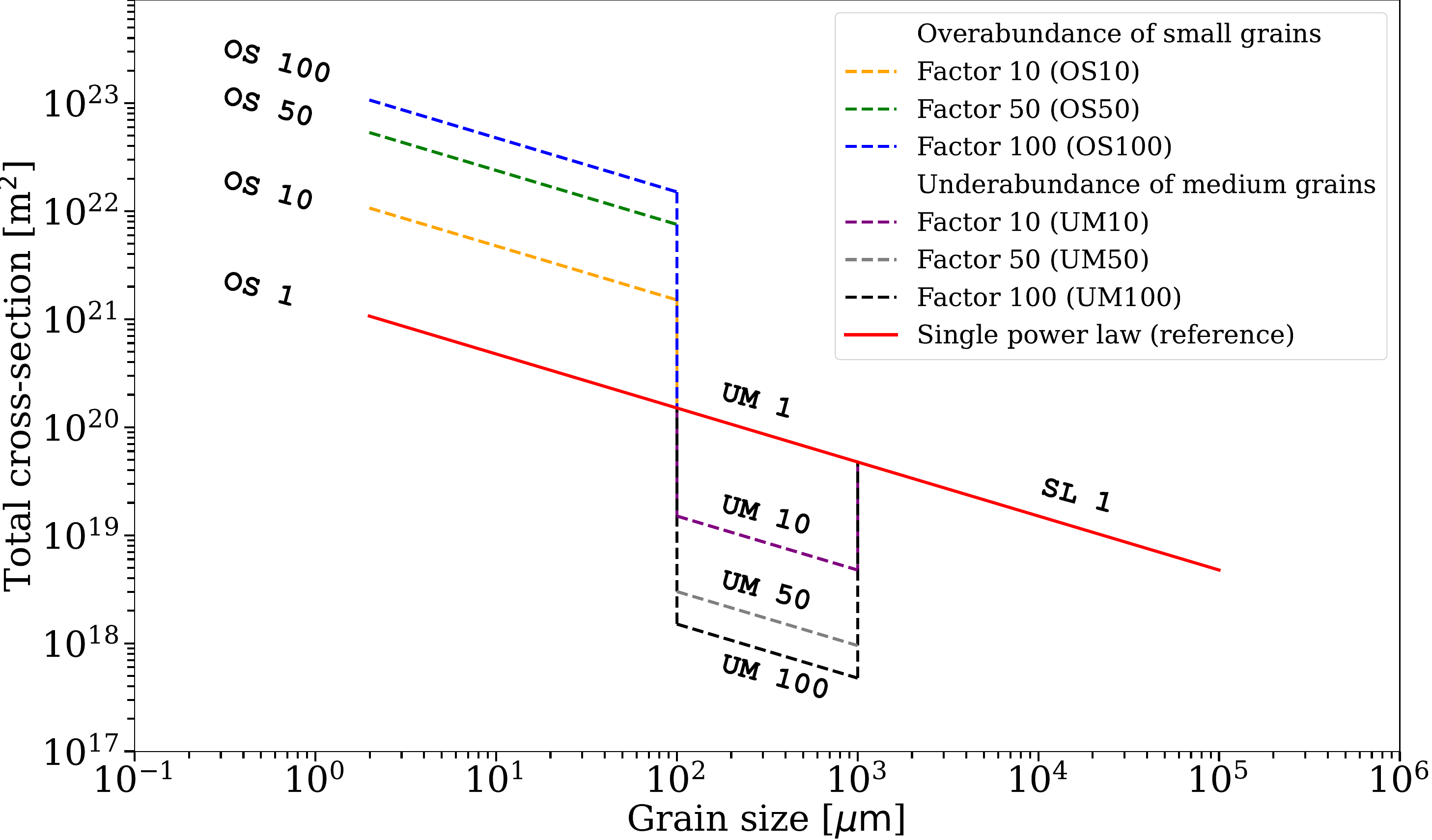}
\caption{Simulated total cross-section of the dust in our debris disk around A star (e.g., $a_{\rm\,bo}$ $\sim$ 1.99 $\mu$m) based on the considered grain size distributions. The red solid line represents the classical single power law (reference case; see Sect.~\ref{Single power law GD}). The yellow, green, and blue dashed lines represent the perturbed size distribution with an overabundance of grains with radii ranging from the blow-out size (see Table~\ref{Star table}) to 100 $\mu{\rm{m}}$. The purple, grey, and navy dashed lines represent the distribution with the underabundance in the size range of 100 $\mu{\rm{m}}$ to 1000 $\mu{\rm{m}}$, which are mainly motivated by the collisional model presented in \citet{Thebault_Augereau_2007} and \citet{Kim2018}.}
\label{fig:GDs}
\end{figure}

As a reference case, we consider the classical single power law grain size distribution d$n\,(a) \propto a^{\,-3.5}$ d$a$ (see Fig.~\ref{fig:GDs} and Table~\ref{Table:grain size distributions} top row for details), resulting from a collisional cascade in debris disks at steady-state equilibrium with size-independent collisional strength (\citealp{Dohnanyi1969}). Since the absorption cross-section $C_{\rm abs}$ of dust grains gradually decreases towards longer wavelengths irrespective of the dust composition (\citealp{Kim2019}), the dust emission model employing the single power law grain size distribution is expected to result in a continuous decrease in the SED at long wavelengths, i.e., the far-IR to mm range.

\begin{table*}[]
\def\arraystretch{1.3}   
\caption{Considered grain size distributions, dust mass, and corresponding model ID ($a_{\rm{bo}}$: blow-out grain size, see Sect.~\ref{Debris disk model parameter} for discussion).}
\centering
\begin{tabular}{llccc}
\hline\hline
Grain size distribution & Description & Factor & Mass of dust [M$_{\,\odot}$] & Model ID\\
\hline
\multirow{2}{*}{Single power law} & Reference model standard abundance & & \multirow{2}{*}{1.00 $\times$ 10$^{-7}$} & \multirow{2}{*}{\texttt{OS1 + UM1 + SL1}}\\ 
& d$n\,(a)$ $\propto$ $a^{\rm - 3.5}$d$a$ for [$a_{\rm\,bo}$ or 0.264 $\mu\rm{m}$, 10 cm] \\\hline
\multirow{9}{*}{res: Broken power law} & \multirow{3}{*}{Overabundance of small grains} & 1   & 2.62 $\times$ 10$^{-10}$ & \texttt{OS1}\\
& \multirow{3}{*}{d$n\,(a)$ $\propto$ $a^{\rm - 3.5}$d$a$ for [$a_{\rm\,bo}$ or 0.264 $\mu\rm{m}$, 100 $\mu\rm{m}$]} & 10 & 2.62 $\times$ 10$^{-9}$ & \texttt{OS10}\\
& & 50 & 1.31 $\times$ 10$^{-8}$ & \texttt{OS50}\\
& & 100 & 2.62 $\times$ 10$^{-8}$ & \texttt{OS100 }\\\cmidrule{2-5}
& \multirow{3}{*}{Underabundance of medium grains}& 1 & 6.65 $\times$ $10^{-9}$ & \texttt{UM1}\\
& \multirow{3}{*}{d$n\,(a)$ $\propto$ $a^{\rm - 3.5}$d$a$ for [100 $\mu\rm{m}$, 1000 $\mu\rm{m}$]}  & 10 & 6.65 $\times$ 10$^{-10}$ & \texttt{UM10}\\
& & 50 & 3.33 $\times$ 10$^{-11}$ & \texttt{UM50}\\
&  & 100 & 6.65 $\times$ 10$^{-11}$ & \texttt{UM100}\\\cmidrule{2-5}
& Standard abundance of large grains & & \multirow{2}{*}{1.00 $\times$ 10$^{-7}$} & \multirow{2}{*}{\texttt{SL1}}\\ 
& d$n\,(a)$ $\propto$ $a^{\rm - 3.5}$d$a$ for [1000 $\mu\rm{m}$, 10 cm] \\\hline
\end{tabular}
\label{Table:grain size distributions}
\end{table*}

\subsubsection{Broken power law grain size distribution}\label{Broken power law GD}

As indicated by previous studies (\citealp{Thebault_Augereau_2007, Loehne2017, Kim2018}), the collisional evolution leaves an imprint on the grain size distribution. In particular, the depletion of the smallest grains to the blow-out size results in a surplus of slightly larger grains due to reduced collisional erosion. This leads to the propagation of an overabundance around small dust grains with radii ranging from several to $\sim$ hundred micrometers, and an underabundance (i.e., local minimum) towards larger grains with radii of several hundred to thousand micrometers. For example, an observed peak in the wavy pattern of the grain size distribution occurs for grains that are slightly larger than the blow-out size, specifically at about $\sim$ 1.5 times this size. Additionally, the first minimum in this distribution is noted at about 100 times the blow-out size (\citealp{Thebault_Augereau_2007}).

Furthermore, for the particular case of debris disks around HD107146, referenced in Sect.~\ref{introduction}, a local decrease in the long-wavelength regime of the SED ranging from 800 $\mu$m to 6 mm has been observed (\citealp{Lestrade2020}). A possible explanation for this finding is to link it to the peculiarities of the grain size distribution, where the deep depletion of the abundance of grains in the hundreds of micrometers size range coincides with the characteristic grain size for strongly absorbing materials~($\sim$ approximately ranging from 100 to 1000~$\mu\rm{m}$; \citealp{Backman1993}).


\begin{figure}
\centerline\centering\includegraphics[width=9cm, height=5.7cm]{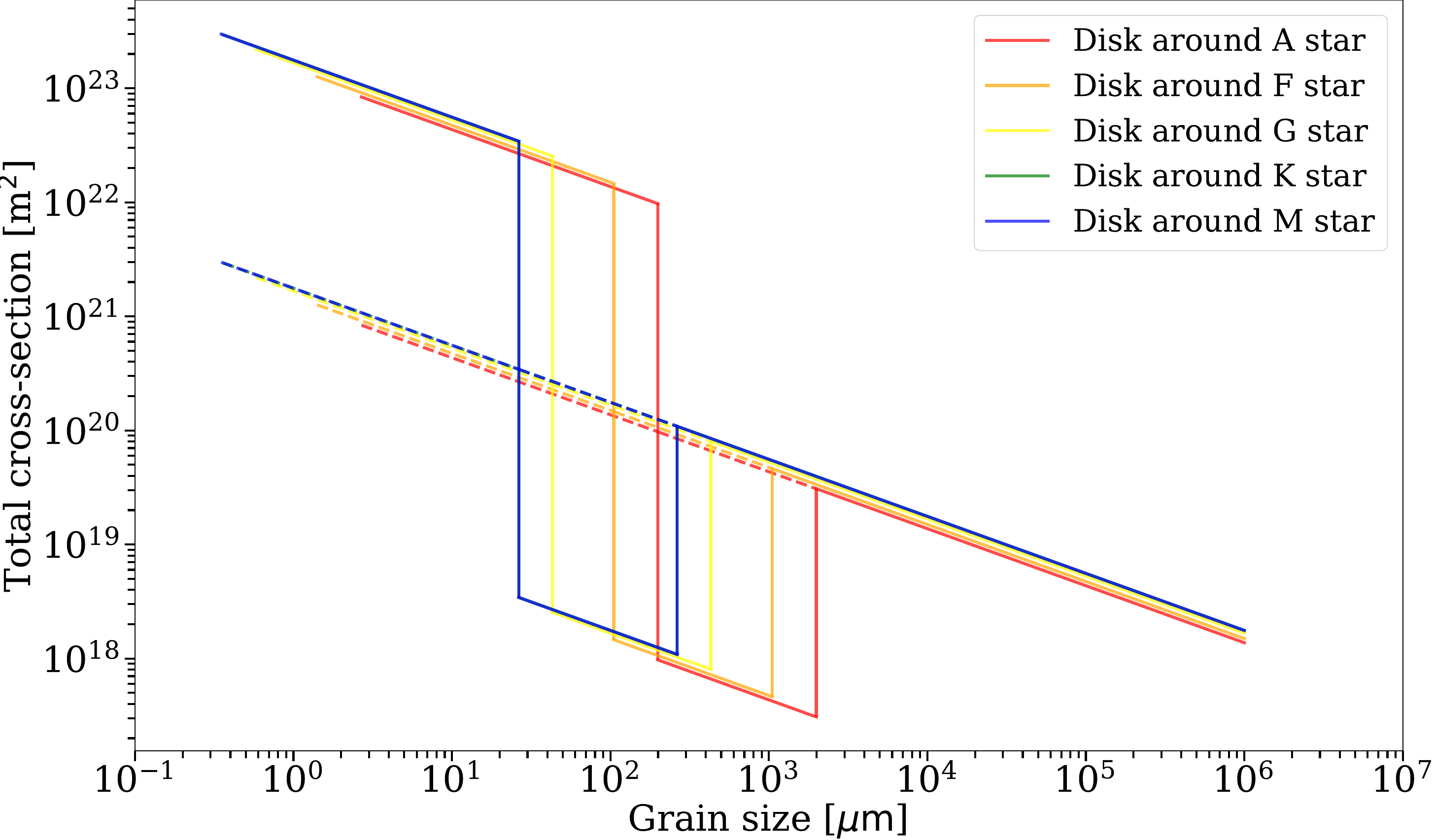}
\caption{Simulated total cross-section of the dust in our debris disk around A, F, G, K, and M star (e.g., $a_{\rm\,bo}$ $\sim$ 1.99, 1.05, 0.43, 0.264, and 0.264, $\mu$m; red, orange, yellow, green, and blue color solid lines) based on the considered grain size distributions with exemplary cases (i.e., 100 times overabundance of small grains and underabundance of medium grains). The dashed red, orange, yellow, green, and blue color lines represent the classical single power law (reference case) around A, F, G, K, and M stars. The perturbed size distribution with an overabundance of small grains with radii ranging from the blow-out size to 100 times each blow-out size depending on spectral types (see Table~\ref{Star table}). The one with an underabundance of medium grains with radii ranging from 100 times to 1000 times each blow-out size depending on spectral types (see Table~\ref{Star table}). These size distributions are mainly motivated by the collisional model presented in \citet{Thebault_Augereau_2007} and \citet{Kim2018}. Note that K star and M star do not have blow-out sizes, suggesting the minimum grain sizes are used often in previous studies (\citealp{Loehne2017, Kim2018}). Thus, both grain size distributions look the same. See Table~\ref{GD_bo_star} for details.}
\label{fig:bo_GDs}
\end{figure}

\begin{table*}[]
\centering
\def\arraystretch{1.3}                          
\caption{The locations of the first and second maximum/minimum in grain size distribution depending on the central star for the simulations. The chosen parameters are based on the findings from previous studies (e.g., \citealp{Thebault_Augereau_2007, Kim2018}).}
\renewcommand{\arraystretch}{1.3}
\begin{tabular}{lllll}
\hline\hline
SpT  &  Blow-out grain size &  first maximum & first minimum & second maximum \\
  &  [$\mu$m] & [$\mu$m] & [$\mu$m] & [$\mu$m]  \\
\hline
A star ($\beta$ Pictoris)  & 1.99 &  2.985 & 199.9 & 1999 \\
F star (HD 139664) 	  & 1.05 &  1.575 & 105 &  1050 \\
G star (HD 107146) 	  &  0.43 &  0.645 &  43 &  430\\
K star (HD 53143)  	   & (0.264) & 0.396 & 26.4 & 264\\
M star (AU mic)    	  & (0.264) &  0.396 & 26.4 & 264\\
\hline
\end{tabular}
\label{GD_bo_star}
\end{table*}
Based on the discussion in this section, we consider general but rather simplistic grain size distribution that shows the key characteristics potentially allowing us to explain discontinuities in the shape of the thermal reemission SED of debris disks. Our discontinuous grain size distribution
models follow a broken power law pattern, comprising an overabundance of small grains ranging from the blow-out grain size to 100 ~$\mu\rm{m}$ and an underabundance of medium-sized grains (in short "medium grain" in the following) ranging from 100 to 1000~$\mu\rm{m}$ (\citealp{Backman1993}). We incorporate scaling factors of 10, 50, and 100 to represent the relative abundance of these grain populations (see Fig.~\ref{fig:GDs} and Table~\ref{Table:grain size distributions} for details). We do not consider the depletion of the smallest dust grains to the blow-out grain size that results in the excess of slightly larger grains (e.g., \citealp{Kim2018}), as the effect of their absence on the spectral slope of the SED is negligible in the spectral range explored in the present study. To avoid the introduction of additional parameters, we set all slopes described by the exponent of the grain size distribution $\gamma$ in the broken power law grain size distribution to -3.5 (\citealp{Dohnanyi1969}) for all size ranges. We note that the location of the subsequent minima/maxima and their overabundance and underabundance may depend on not only the collisional circumstances but also further disk parameters (e.g., \citealp{Thebault_Augereau_2007, Kim2018}). Thus, the location and abundance of the local minima/maxima of the size distribution may not be directly indicated by the physical situations (e.g., the collisional state of the disk). \newline 

\noindent We note that a uniform overabundance and/or underabundance of grains of certain sizes across all spectral types may not accurately represent the actual conditions within these systems. Thus, we particularly consider grain size distributions with an overabundance and/or underabundance of certain grain sizes as a function of stellar spectral type (i.e., blow-out size). For this purpose, we apply a grain size distribution model following the findings of \citealp{Thebault_Augereau_2007, Kim2018}, in which the first local maximum of the grain size distribution typically occurs at about $\sim$ 1.5 times the blow-out size. Furthermore, the first minimum and second maximum are observed at around 100 and 1000 times the blow-out size, respectively (see also Fig.~\ref{fig:bo_GDs} and Table~\ref{GD_bo_star}). In the case of disks around stars that are not luminous enough to reach $\beta$ = 0.5 (e.g., K and M stars in our model; \citealp{Kirchschlager2013}), we consider 0.264 $\mu$m as the minimum grain size (see also Sect.~\ref{Debris disk model parameter}). The results based on the specifically proposed grain size distribution, particularly concerning its impact on the corresponding SEDs, can be found in Sect.~\ref{Effect of the blow-out size dependent grain size distribution on the SEDs}.

\section{Results}\label{Results}

In the following, we systematically compare and evaluate the influence of the different dust emission models based on the two different grain size distributions on the SED for the various considered debris disk systems. Specifically, we focus on quantifying the slope of the SED at long wavelengths to assess how the presence of over- and underabundances of the specific size of grains in the distribution impacts the emission characteristics of debris disks. For this purpose, we compute the spectral index ${\alpha}$ as,
\begin{equation}
{\alpha} = \left|{\dfrac{{\rm{log}}\,({F_{\nu 1}}/{F_{\nu 2}})}{{\rm{log}}\,({\nu_{1}}/{\nu_{2}})}}\right|,
\label{eqn: spectral index}
\end{equation}

\noindent where $F_{\nu}$ represents the radiative flux density and $\nu$ denotes the frequency in the dust continuum emission.

\begin{table*}[]
\small
\centering
\def\arraystretch{1.5}   
\caption{Comparison of the spectral index of the considered discontinuous grain size distributions depending on the overabundance of small grains, stellar parameters (e.g., stellar temperature and radius), and disk parameters (e.g., fractional width $\Delta$$R$/$R$). For the definition of model IDs (\texttt{OS, UM}, and \texttt{SL}) see Table~\ref{Table:grain size distributions}.}
\begin{tabular}{llccc|cccccccccc}
\hline\hline
\multirow{3}{*}{SpT} & \multirow{3}{*}{$\Delta$$R$/$R$} & \multicolumn{3}{c}{Single power law} & \multicolumn{9}{c}{overabundance of small grains} \\\cmidrule{3-14}
& & \multicolumn{3}{c}{\texttt{OS1+UM1+SL1}}& \multicolumn{3}{c}{\texttt{OS10+UM1+SL1}} & \multicolumn{3}{c}{\texttt{OS50+UM1+SL1}}& \multicolumn{3}{c}{\texttt{OS100+UM1+SL1}}\\
\cmidrule{3-14}
&  & ${\alpha}^{\,450\,\mu{\rm{m}}}_{\,850\,\mu{\rm{m}}}$ & ${\alpha}^{\,1000\,\mu{\rm{m}}}_{\,2000\,\mu{\rm{m}}}$ & $\Delta\,\alpha$ &${\alpha}^{\,450\,\mu{\rm{m}}}_{\,850\,\mu{\rm{m}}}$ & ${\alpha}^{\,1000\,\mu{\rm{m}}}_{\,2000\,\mu{\rm{m}}}$ & $\Delta\,\alpha$ & ${\alpha}^{\,450\,\mu{\rm{m}}}_{\,850\,\mu{\rm{m}}}$ & ${\alpha}^{\,1000\,\mu{\rm{m}}}_{\,2000\,\mu{\rm{m}}}$ & $\Delta\,\alpha$ & ${\alpha}^{\,450\,\mu{\rm{m}}}_{\,850\,\mu{\rm{m}}}$ & ${\alpha}^{\,1000\,\mu{\rm{m}}}_{\,2000\,\mu{\rm{m}}}$ & $\Delta\,\alpha$  \\\hline
\multirow{2}{*}{A}   & 0.17  & 2.07  & 2.75 & -0.68 & 2.59      & 2.77      & -0.18    & 3.67      & 2.86      & 0.81     & 4.24       & 2.95      & 1.28     \\
                     & 1.25  &2.08  & 2.75 & -0.67 & 2.6       & 2.78      & -0.18    & 3.68      & 2.87      & 0.81     & 4.25       & 2.96      & 1.29     \\\hline
\multirow{2}{*}{F}   & 0.17 & 1.99  & 2.76 & -0.77 & 2.51      & 2.79      & -0.28    & 3.58      & 2.88      & 0.7      & 4.13       & 2.98      & 1.15     \\
                     & 1.25 &2.00  & 2.77 & -0.77 & 2.52      & 2.79      & -0.27    & 3.59      & 2.88      & 0.71     & 4.14       & 2.98      & 1.16     \\\hline
\multirow{2}{*}{G}   & 0.17 & 1.93  & 2.76 & -0.83 & 2.44      & 2.79      & -0.35    & 3.49      & 2.88      & 0.61     & 4.03       & 2.98      & 1.05     \\
                     & 1.25 &1.94  & 2.77 & -0.83 & 2.45      & 2.79      & -0.34    & 3.51      & 2.89      & 0.62     & 4.05       & 2.98      & 1.06     \\\hline
\multirow{2}{*}{K}   & 0.17 & 1.86  & 2.75 & -0.89  & 2.36      & 2.77      & -0.41    & 3.41      & 2.87      & 0.54     & 3.94       & 2.97      & 0.97     \\
                     & 1.25 &1.88  & 2.75 & -0.88 & 2.38      & 2.78      & -0.4     & 3.43      & 2.87      & 0.56     & 3.96       & 2.97      & 0.98     \\\hline
\multirow{2}{*}{M}   & 0.17 & 1.53  & 2.64 & -1.11 & 2.06      & 2.67      & -0.61    & 3.12      & 2.78      & 0.34     & 3.63       & 2.89      & 0.74     \\
                     & 1.25 &1.57  & 2.65 & -1.08 & 2.1       & 2.68      & -0.58    & 3.15      & 2.79      & 0.36     & 3.67       & 2.89      & 0.77    \\\hline
\end{tabular}
\label{table:SED with broken power law grain size distribution: overabundance of small grains}
\end{table*}

\begin{figure*}
\centerline\centering\includegraphics[width=18cm, height=12cm]{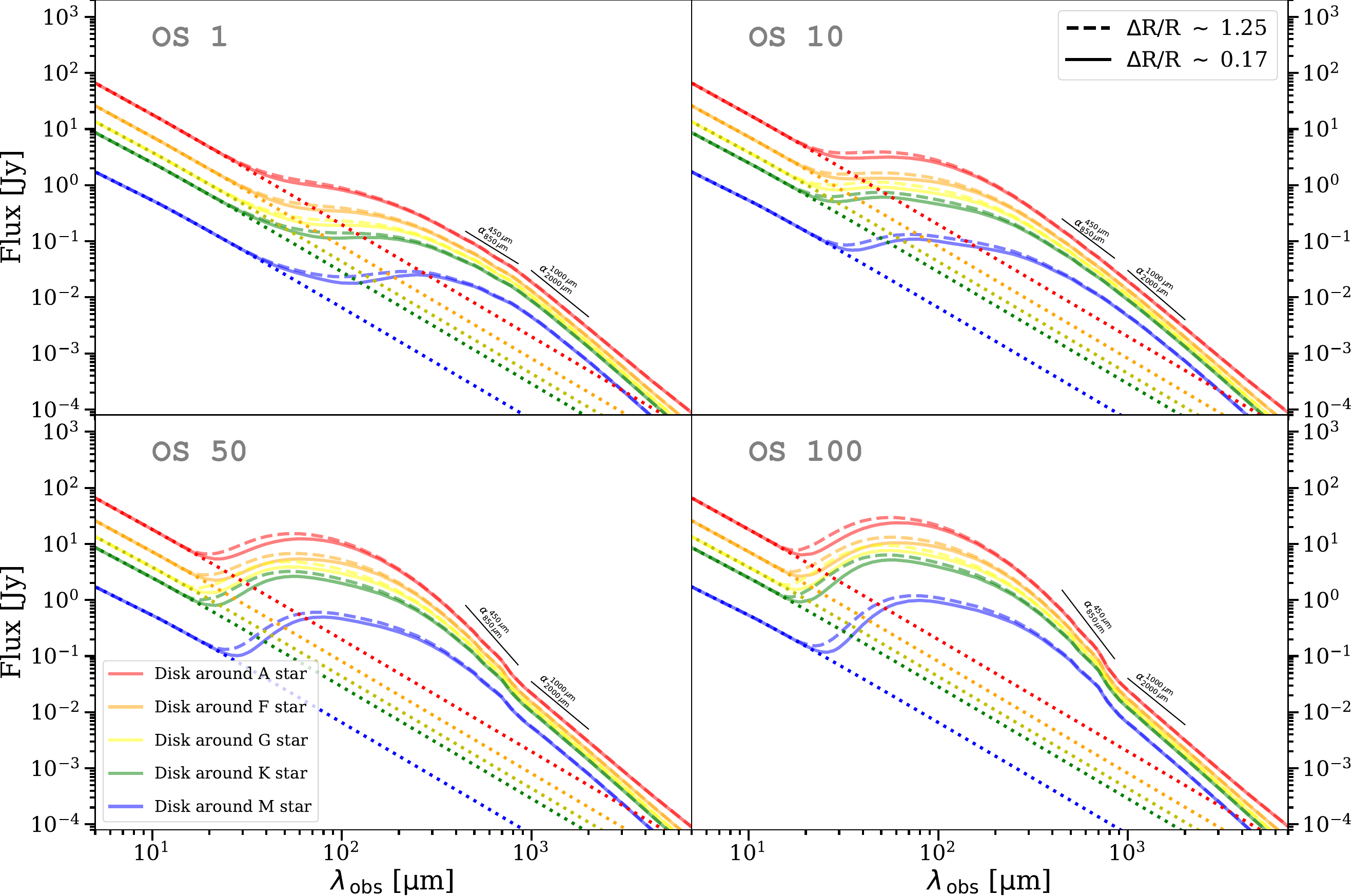}
\caption{SEDs of the case of the overabundance of small grains. Solid and dashed lines indicate the modeled SED of the dust reemission for the two different disk structures (thin belts and broad disks). The dotted lines represent the direct stellar radiation. The top left plot shows the result based on the single power law grain size distribution (\texttt{OS1+UM1+SL1}), while the other three plots exhibit the results based on the broken power law grain size distribution, specifically the overabundance of small grains with the different scaling factors, 10, 50, and 100 (i.e., \texttt{OS10}, \texttt{OS50}, and \texttt{OS100}). See Sect.~\ref{res: Broken power law} and Table~\ref{table:SED with broken power law grain size distribution: overabundance of small grains} for details.}
\label{fig:SED with broken power law grain size distribution: overabundance of small grains}
\end{figure*}

Considering our specific grain size distribution with overabundant small grains and a deep depletion of grains in the hundreds of micrometers size range, we expect spectral transitions to occur at sub-mm to mm wavelengths. To characterize the SED in this wavelength range, we calculate the spectral index within two distinct wavelength ranges: 450/850 $\mu{\rm{m}}$ (${\alpha}^{\,450\,\mu{\rm{m}}}_{\,850\,\mu{\rm{m}}}$) and 1000/2000 $\mu{\rm{m}}$ (${\alpha}^{\,1000\,\mu{\rm{m}}}_{\,2000\,\mu{\rm{m}}}$). These spectral indices enable us to quantify the differences between the two wavelength ranges, represented by $\Delta\,\alpha$ = ${\alpha}^{\,450\,\mu{\rm{m}}}_{\,850\,\mu{\rm{m}}}$ - ${\alpha}^{\,1000\,\mu{\rm{m}}}_{\,2000\,\mu{\rm{m}}}$. Positive and negative values of $\Delta\,\alpha$ indicate a lower and higher value of ${\alpha}^{\,1000\,\mu{\rm{m}}}_{\,2000\,\mu{\rm{m}}}$, representing a shallower and steeper slope in 1000 to 2000 $\mu{\rm{m}}$ range compared to 450 to 850 $\mu{\rm{m}}$ range, respectively.

\subsection{Single power law grain size distribution}\label{res: Single power law}

We begin by examining the spectral index of the SED for the continuous grain size distribution described by the single power law (i.e., $\gamma$ = -3.5). This reference case will allow us to identify the influence of the over/underabundance of grains of specific sizes. The top left plot of Figs. \ref{fig:SED with broken power law grain size distribution: overabundance of small grains} and \ref{fig:SED with broken power law grain size distribution: underabundance of mid grains} shows the resulting SED for the two different disk structures (i.e., thin belt and broad disk) depending on the spectral type of the central stars (see also Tables \ref{table:SED with broken power law grain size distribution: overabundance of small grains} and \ref{table:SED with broken power law grain size distribution: underabundance of mid grains}, column "\texttt{OS1}+\texttt{UM1}+\texttt{SL1}"). For this reference case, the SEDs show the "typical" smooth and steady decrease in the FIR/mm wavelength range, resulting from the steady decrease of the absorption cross-section $C_{\rm{abs}}$ towards longer wavelengths and the steady decrease of the dust grain temperature with increasing radial distance from the central star. 

The slope of the SED slightly steepens with increasing the wavelength, i.e., the value of ${\alpha}^{\,1000\,\mu{\rm{m}}}_{\,2000\,\mu{\rm{m}}}$ is slightly larger than ${\alpha}^{\,450\,\mu{\rm{m}}}_{\,850\,\mu{\rm{m}}}$. The absolute amount of this difference between the spectral indices increases towards stars of late spectral type. This can be explained by the shift of the wavelength of the dust emission maximum towards longer wavelengths with decreasing effective temperature of the central star. Consequently, the spectral index ${\alpha}^{\,450\,\mu{\rm{m}}}_{\,850\,\mu{\rm{m}}}$ decreases, leading to an increase of the absolute value of $\Delta\,\alpha$.


\subsection{Broken power law grain size distribution}\label{res: Broken power law}
We now analyze the spectral index of the SED for models with a discontinuous grain size distribution described by a broken power law. Our grain size distribution model is defined by three distinct groups of grains: an overabundance of small grains, a depletion of medium-sized grains, and a standard abundance of larger grains (Sect.~\ref{Broken power law GD}). We assess the individual impact of each group on the resulting SEDs separately.\newline

\noindent \textbf{Overabundance of small grains} \hspace{2mm} Fig.~\ref{fig:SED with broken power law grain size distribution: overabundance of small grains} shows the modeled SED with a broken power law grain size distribution depending on the overabundance of small grains, disk structures (thin belts and broad disks), and varying stellar sources. The corresponding spectral indices ${\alpha}^{\,450\,\mu{\rm{m}}}_{\,850\,\mu{\rm{m}}}$, ${\alpha}^{\,1000\,\mu{\rm{m}}}_{\,2000\,\mu{\rm{m}}}$, and their differences $\Delta\,\alpha$ are listed in Table~\ref{table:SED with broken power law grain size distribution: overabundance of small grains}.

\begin{figure}
\centerline\centering\includegraphics[width=9cm, height=6cm]{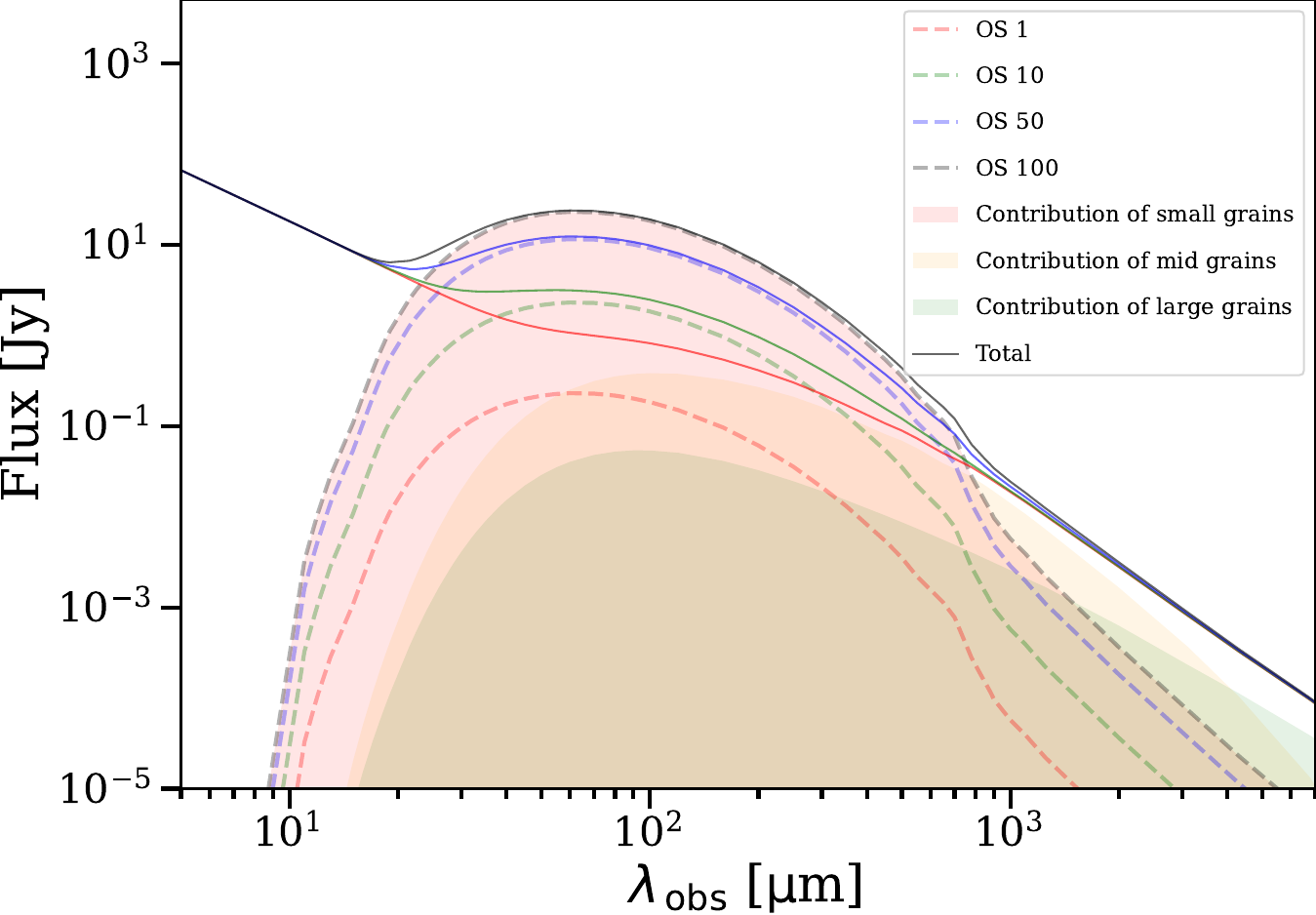}
\caption{Modeled SED of debris disks with a narrow belt ($\Delta$$R$/$R$ $\sim$ 0.17) around an A-type star depending on the overabundance of small grains (color dot lines). Total emissions (i.e., net SED including the emission from the star and dust) are indicated with solid lines of the same color for the overabundant of small grains (dashed line). See Table~\ref{table:SED with broken power law grain size distribution: overabundance of small grains} for details.}
\label{fig:SED contribution from small size grains}
\end{figure}

We find that an overabundance of small grains leads to a modest impact on the mm slope of the SED, causing it to become slightly steeper (i.e., slightly higher value of ${\alpha}^{\,1000\,\mu{\rm{m}}}_{\,2000\,\mu{\rm{m}}}$), but predominantly affects and steepens the far-IR slope of the SED (i.e., higher value of ${\alpha}^{\,450\,\mu{\rm{m}}}_{\,850\,\mu{\rm{m}}}$). This behavior becomes more pronounced as the overabundance of small grains increases. In particular, for models with an overabundance of small grains of a factor of 50 (\texttt{OS50}) -- compared to the ideal collisional cascade model -- we find a significant kink in the slope of the SED in the 800 to 900 $\mu$m wavelength range (see the bottom left plot of Fig.~\ref{fig:SED with broken power law grain size distribution: overabundance of small grains}). 


Furthermore, we find that the spectral index transition is more pronounced in debris disks around brighter (hotter) central stars and larger radial extent. The systems with broader debris rings (i.e., disks with an inner radius that is close significantly close to the central star and an outer radius that is located farther out, compared to their thin-ring counterpart) exhibit a larger range of temperatures, with higher dust temperatures closer to the star, resulting in stronger emissions with the location of their maximum shifted towards shorter wavelengths. This leads to a steeper slope of the SED in the 450 - 850 $\mu\rm{m}$ wavelength range, e.g., larger absolute values of the index ${\alpha}^{\,450\,\mu{\rm{m}}}_{\,850\,\mu{\rm{m}}}$. However, compared to the impact of different abundance in the different grain size distribution, the impact of the fractional width $\Delta\,R$/$R$ on the spectral index is small.

Next, we investigate the contribution of grains in different size regimes to the observed spectral index. For this purpose, we examine their individual contributions to the total flux and thus their impact on the spectral indices (see
Fig.~\ref{fig:SED contribution from small size grains}). Dust particles with sizes comparable to the observing wavelength emit most efficiently in this wavelength regime (\citealp{Backman1993}). We find that small grains, which constitute a significant portion of the emitting surface area in the grain size distribution, primarily affect the spectral index ${\alpha}^{\,450\,\mu{\rm{m}}}_{\,850\,\mu{\rm{m}}}$ as they efficiently re-emit in the far-IR, while their contribution to the long wavelength regime, and thus their impact on ${\alpha}^{\,1000\,\mu{\rm{m}}}_{\,2000\,\mu{\rm{m}}}$, is negligible. On the other hand, grains larger than the medium size (e.g., \texttt{SL1}), whose abundance is held constant in this model, have a noticeable impact on the SED and the spectral index in the mm range. In particular, grains larger than $\sim$ 1 mm, which radiate like black bodies at longer wavelengths, cause a shallower slope of the mm SED. \newline

\begin{figure*}
\centerline\centering\includegraphics[width=18cm, height=12cm]{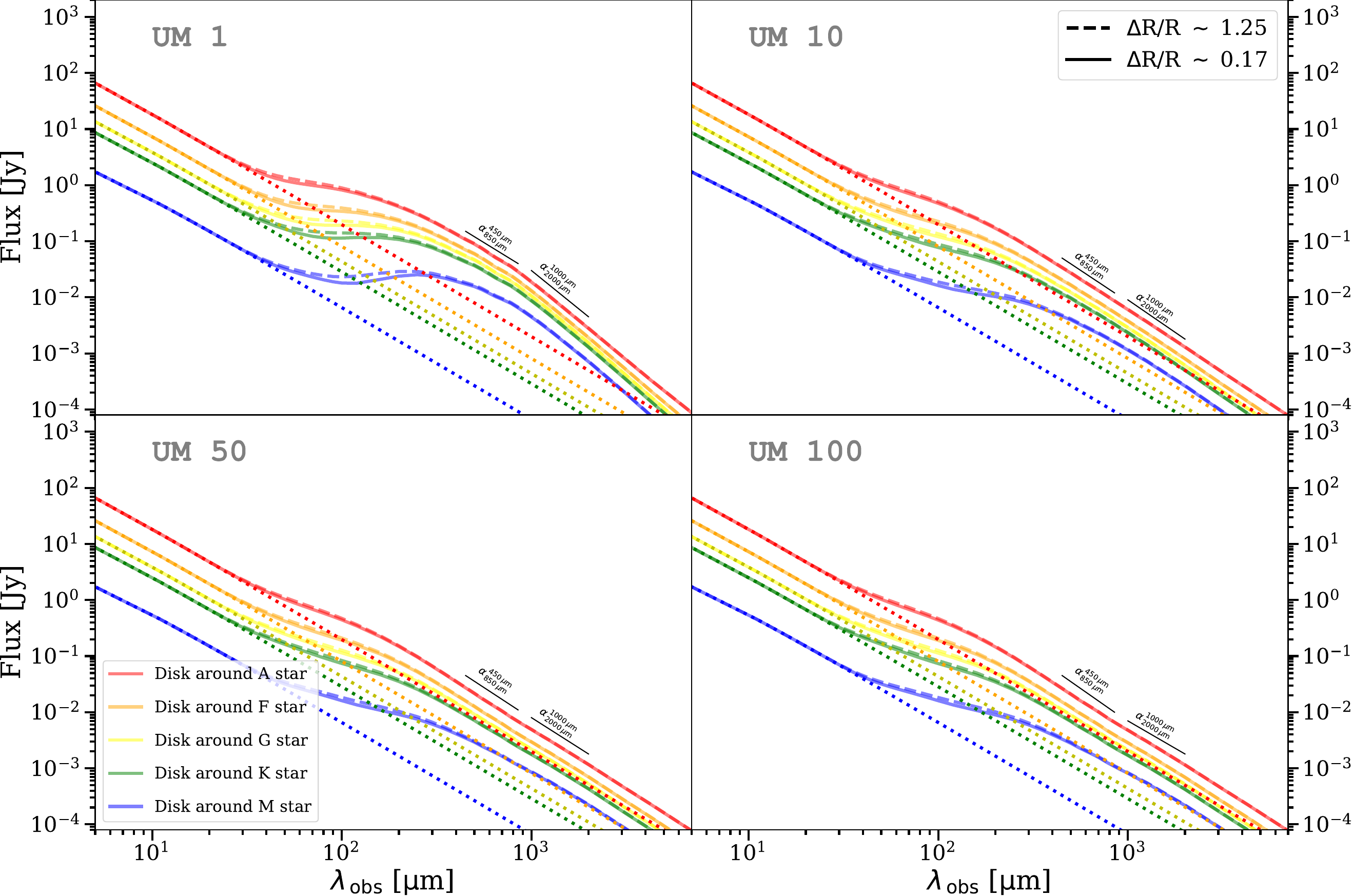}
\caption{SEDs of the case of the underabundance of medium grains. Solid and dashed lines indicate the modeled SED of the dust reemission for the two different disk structures (thin belts and broad disks). The dotted lines represent the direct stellar radiation. The top left plot shows the result based on the single power law grain size distribution (\texttt{OS1+UM1+SL1}), while the other three plots exhibit the results based on the broken power law grain size distribution, specifically the underabundance of medium grains with the different scaling factors, 10, 50, and 100 (i.e., \texttt{UM10}, \texttt{UM50}, and \texttt{UM100}). See Sect.~\ref{res: Broken power law} and Table~\ref{table:SED with broken power law grain size distribution: underabundance of mid grains} for details.}
\label{fig:SED with broken power law grain size distribution: underabundance of mid grains}
\end{figure*}

\noindent \textbf{Underabundance of medium grains} \hspace{2mm} Fig.~\ref{fig:SED with broken power law grain size distribution: underabundance of mid grains} shows the SED for the case of the broken power law grain size distribution depending on the underabundance of medium grains, disk structures (thin belts and broad disks), and spectral type of the central stars. The corresponding spectral indices ${\alpha}^{\,450\,\mu{\rm{m}}}_{\,850\,\mu{\rm{m}}}$, ${\alpha}^{\,1000\,\mu{\rm{m}}}_{\,2000\,\mu{\rm{m}}}$, and their differences $\Delta\,\alpha$ are listed in Table~\ref{table:SED with broken power law grain size distribution: underabundance of mid grains}.

\begin{table*}[]
\centering
\small
\def\arraystretch{1.5}   
\caption{Comparison of the spectral index of the considered discontinuous grain size distributions depending on the underabundance of medium grains, stellar parameters (e.g., stellar temperature and radius), and disk parameters (e.g., fractional width $\Delta$$R$/$R$). For the definition of model IDs (\texttt{OS, UM}, and \texttt{SL}) see Table~\ref{Table:grain size distributions}.}
\begin{tabular}{llccc|cccccccccc}
\hline\hline
\multirow{3}{*}{SpT} & \multirow{3}{*}{$\Delta$$R$/$R$} & \multicolumn{3}{c}{Single power law} & \multicolumn{9}{c}{underabundance of medium grains} \\\cmidrule{3-14}
 &  & \multicolumn{3}{c}{\texttt{OS1+UM1+SL1}}& \multicolumn{3}{c}{\texttt{OS1+UM10+SL1}} & \multicolumn{3}{c}{\texttt{OS1+UM50+SL1}} & \multicolumn{3}{c}{\texttt{OS1+UM100+SL1}} \\
\cmidrule{3-14}
& & ${\alpha}^{\,450\,\mu{\rm{m}}}_{\,850\,\mu{\rm{m}}}$ & ${\alpha}^{\,1000\,\mu{\rm{m}}}_{\,2000\,\mu{\rm{m}}}$ & $\Delta\,\alpha$ & ${\alpha}^{\,450\,\mu{\rm{m}}}_{\,850\,\mu{\rm{m}}}$ & ${\alpha}^{\,1000\,\mu{\rm{m}}}_{\,2000\,\mu{\rm{m}}}$ & $\Delta\,\alpha$ & ${\alpha}^{\,450\,\mu{\rm{m}}}_{\,850\,\mu{\rm{m}}}$ & ${\alpha}^{\,1000\,\mu{\rm{m}}}_{\,2000\,\mu{\rm{m}}}$ & $\Delta\,\alpha$ & ${\alpha}^{\,450\,\mu{\rm{m}}}_{\,850\,\mu{\rm{m}}}$ & ${\alpha}^{\,1000\,\mu{\rm{m}}}_{\,2000\,\mu{\rm{m}}}$ & $\Delta\,\alpha$  \\\hline
\multirow{2}{*}{A}   & 0.17 &2.07  & 2.75 & -0.68  & 2.15      & 2.22      & -0.07    & 2.17      & 2.07      & 0.1      & 2.18       & 2.05      & 0.13     \\
 & 1.25 &2.08  & 2.75 & -0.67   & 2.15      & 2.22      & -0.07    & 2.18      & 2.08      & 0.1      & 2.19       & 2.06      & 0.13     \\\hline
\multirow{2}{*}{F}   & 0.17 & 1.99  & 2.76 & -0.77 & 2.09      & 2.23      & -0.14    & 2.13      & 2.07      & 0.06     & 2.14       & 2.04      & 0.1 \\ 
& 1.25 &2.00  & 2.77 & -0.77   & 2.1       & 2.24      & -0.14    & 2.14      & 2.07      & 0.07     & 2.15       & 2.05      & 0.1      \\\hline
\multirow{2}{*}{G}   & 0.17  &1.93  & 2.76 & -0.83   & 2.04      & 2.23      & -0.2     & 2.08      & 2.06      & 0.03     & 2.09       & 2.03      & 0.06 \\ 
& 1.25  &1.94  & 2.77 & -0.83  & 2.05      & 2.24      & -0.19    & 2.1       & 2.06      & 0.03     & 2.1        & 2.03      & 0.07     \\\hline
\multirow{2}{*}{K}   & 0.17  & 1.86  & 2.75 & -0.89 & 1.97 & 2.22      & -0.25    & 2.03      & 2.04      & -0.01    & 2.04       & 2.01      & 0.02     \\ & 1.25 & 1.88  & 2.75 & -0.88  & 1.99      & 2.23      & -0.24    & 2.04      & 2.04      & 0        & 2.05       & 2.02      & 0.03     \\\hline
\multirow{2}{*}{M}   & 0.17 & 1.53  & 2.64 & -1.11     & 1.68      & 2.13      & -0.44    & 1.75    & 1.94      & -0.19    & 1.76       & 1.91      & -0.14    \\  
& 1.25 &1.57  & 2.65 & -1.08   & 1.71      & 2.13      & -0.42    & 1.78      & 1.94      & -0.16    & 1.8        & 1.91      & -0.12  \\ \hline
\end{tabular}
\label{table:SED with broken power law grain size distribution: underabundance of mid grains}
\end{table*}

\begin{figure}
\centerline\centering\includegraphics[width=9cm, height=6cm]{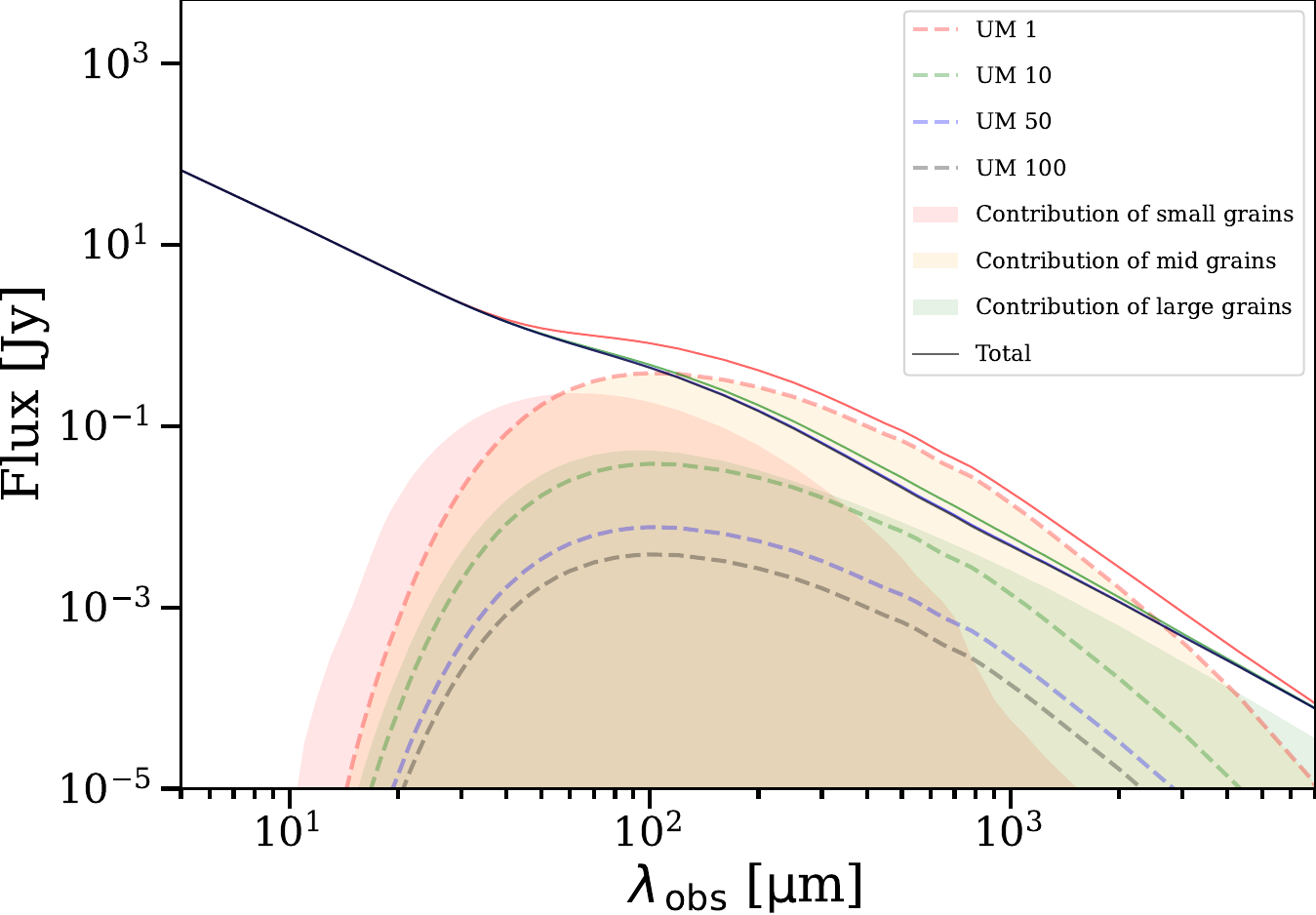}
\caption{Modeled SED of debris disks with a narrow belt ($\Delta$$R$/$R$ $\sim$ 0.17) around an A-type star depending on the underabundance of medium grains (color dot lines). Total emissions (i.e., net SED including the emission from the star and dust) are indicated with solid lines of the same color for the underabundance of medium grains. See Table~\ref{table:SED with broken power law grain size distribution: overabundance of small grains} for details.}
\label{fig:SED contribution from mid size grains}
\end{figure}

For an exemplary case, the contribution of medium grains for different levels of their underabundance to the SED is shown in Fig.~\ref{fig:SED contribution from mid size grains}. We find that an underabundance of medium grains leads to a slight steepening of the far-IR slope of the SED (i.e., the slightly higher value of ${\alpha}^{\,450\,\mu{\rm{m}}}_{\,850\,\mu{\rm{m}}}$). However, it primarily affects the mm slope of the SED, which becomes shallower, corresponding to a lower spectral index ${\alpha}^{\,1000\,\mu{\rm{m}}}_{\,2000\,\mu{\rm{m}}}$. Therefore, the presence of both an overabundance of small grains and an underabundance of medium grains increases the absolute difference between ${\alpha}^{\,450\,\mu{\rm{m}}}_{\,850\,\mu{\rm{m}}}$ and ${\alpha}^{\,1000\,\mu{\rm{m}}}_{\,2000\,\mu{\rm{m}}}$, leaving the 800 - 900 $\mu$m kink as a characteristic impact on the reemission SED. 

However, we do not observe a significant spectral inversion solely due to the underabundance of medium grains. Thus, we conclude that the impact of the underabundance of medium grains on the spectral transition is weaker compared to the effect of the overabundance of small grains. Additionally, we find that the spectral index transition is more pronounced in debris disks around brighter (hotter) central stars and broader disk systems, which is consistent with the findings regarding the overabundance of small grains.

We note that the impact of the underabundance of medium grains could lead to the higher value of ${\alpha}^{\,450\,\mu{\rm{m}}}_{\,850\,\mu{\rm{m}}}$ if the disk is massive enough (e.g., more mass in the medium to large-sized particles). However, this model could also lead to a steeper slope of the SED at a longer wavelength (i.e., the higher value of ${\alpha}^{\,1000\,\mu{\rm{m}}}_{\,2000\,\mu{\rm{m}}}$). Thus, the grain size model with the underabundance of medium grains hardly shows a  noticeable spectral inversion. \newline

\section{Discussion}\label{Discussion}

\subsection{Application to observations}\label{Dis: Application to observations}
\citet{Lestrade2020} highlighted that the spectral indices ${\alpha}^{\,1153\,\mu{\rm{m}}}_{\,2000\,\mu{\rm{m}}}$ derived from NIKA2 observations of the debris disks around Vega and HD 107146 are in agreement with the expected values in the Rayleigh-Jeans regime, consistently around 2.0 $\pm$ 0.18 to 2.0 $\pm$ 0.32. Furthermore, \citealt{MacGregor2016} found that mm spectral indices ${\alpha}_{\rm mm}$ (i.e., the flux density for the complete sample at a given mm wavelength combined with the VLA 9 mm or ATCA 7 mm fluxes; \citealp{Ricci2012, Ricci2015a, Ricci2015b}) have the value of 1.63 to 3.08 with mean values $\sim$ 2.55, inferring for the slope of the size distribution a range from -2.84 (e.g., HD 141569) to -3.64 (e.g., HD 104860) with the weighted mean of these value of -3.36 $\pm$ 0.02. These results indicate a shallower slope of size distribution compared to standard collisional models. \citet{Matthews2015} also indicated a spectral index of $\sim$ 2 desired at sub-mm/mm wavelengths for AU Mic, which is lower than expected. These findings are in contrast to the steeper spectral indices ${\alpha}^{\,450\,\mu{\rm{m}}}_{\,850\,\mu{\rm{m}}}$ measured at shorter wavelengths, showing 2.8 to 2.9 (\citealp{Holland2017}). In general, the SONS JCMT/SCUBA2 Legacy survey (\citealp{Holland2017}) found that the spectral indices ${\alpha}^{\,450\,\mu{\rm{m}}}_{\,850\,\mu{\rm{m}}}$ in 48 debris disks range from 2.7 to 4.7, indicating some size distributions steeper than one based on the collisional cascade.

Further spectral index inversions were observed in the debris disks around q1 Eri (i.e., from 2.90 for ${\alpha}^{\,350\,\mu{\rm{m}}}_{\,860\,\mu{\rm{m}}}$ to 2.39 for ${\alpha}^{\,1260\,\mu{\rm{m}}}_{\,6800\,\mu{\rm{m}}}$; \citealp{Holland2017}, \citealp{Ricci2015b}), HD 98800 (i.e., from 3.21 for ${\alpha}^{\,350\,\mu{\rm{m}}}_{\,870\,\mu{\rm{m}}}$ to 1.00 for ${\alpha}^{\,1350\,\mu{\rm{m}}}_{\,2000\,\mu{\rm{m}}}$; \citealp{Nilsson2010}, \citealp{Sylvester2001}), and HD 95086 (i.e., from 3.18 for ${\alpha}^{\,500\,\mu{\rm{m}}}_{\,1300\,\mu{\rm{m}}}$ to 2.34 for ${\alpha}^{\,1300\,\mu{\rm{m}}}_{\,6800\,\mu{\rm{m}}}$; \citealt{Su2017}, \citealp{Ricci2015b}).

\subsection{Possible degeneracies with the grain composition}\label{Dis: Possible degeneracies with the grain composition}

\begin{table*}[]
\small
\centering
\def\arraystretch{1.5}   
\caption{Comparison of the spectral index of the considered discontinuous grain size distributions depending on the overabundance of small grains, disk parameters (e.g., fractional width $\Delta$$R$/$R$), and ice volume fraction $\mathcal{F_{\rm{ice}}}$. For the definition of model IDs (\texttt{OS, UM}, and \texttt{SL}) see Table~\ref{Table:grain size distributions}.}
\begin{tabular}{llccc|cccccccccc}
\hline\hline
\multirow{3}{*}{Composition} & \multirow{3}{*}{$\Delta$$R$/$R$} & \multicolumn{3}{c}{Single power law} & \multicolumn{9}{c}{overabundance of small grains (\texttt{OS}) or underabundance of medium grains (\texttt{UM}) } \\\cmidrule{3-14}
& & \multicolumn{3}{c}{\texttt{OS1+UM1+SL1}}& \multicolumn{3}{c}{\texttt{OS10+UM1+SL1}} & \multicolumn{3}{c}{\texttt{OS50+UM1+SL1}}& \multicolumn{3}{c}{\texttt{OS100+UM1+SL1}}\\
\cmidrule{3-14}
&  & ${\alpha}^{\,450\,\mu{\rm{m}}}_{\,850\,\mu{\rm{m}}}$ & ${\alpha}^{\,1000\,\mu{\rm{m}}}_{\,2000\,\mu{\rm{m}}}$ & $\Delta\,\alpha$ &${\alpha}^{\,450\,\mu{\rm{m}}}_{\,850\,\mu{\rm{m}}}$ & ${\alpha}^{\,1000\,\mu{\rm{m}}}_{\,2000\,\mu{\rm{m}}}$ & $\Delta\,\alpha$ & ${\alpha}^{\,450\,\mu{\rm{m}}}_{\,850\,\mu{\rm{m}}}$ & ${\alpha}^{\,1000\,\mu{\rm{m}}}_{\,2000\,\mu{\rm{m}}}$ & $\Delta\,\alpha$ & ${\alpha}^{\,450\,\mu{\rm{m}}}_{\,850\,\mu{\rm{m}}}$ & ${\alpha}^{\,1000\,\mu{\rm{m}}}_{\,2000\,\mu{\rm{m}}}$ & $\Delta\,\alpha$  \\\hline
\multirow{2}{*}{Astrosil}   & 0.17  & 2.07  & 2.75 & -0.68 & 2.59      & 2.77      & -0.18    & 3.67      & 2.86      & 0.81     & 4.24       & 2.95      & 1.28     \\
 & 1.25  &2.08  & 2.75 & -0.67 & 2.6       & 2.78      & -0.18    & 3.68      & 2.87      & 0.81     & 4.25       & 2.96      & 1.29     \\\hline
\multirow{2}{*}{$\mathcal{F_{\rm{ice}}}$ = 0.5}   & 0.17  & 2.55 & 2.82 & -0.26 & 2.92 & 2.85 & 0.07    & 3.72      & 2.99      & 0.73     & 4.12       & 3.11      & 1.01     \\
                     & 1.25  &2.56 & 2.82 & -0.26 & 2.93 & 2.86 & 0.07    & 3.73 & 3.00 & 0.73     & 4.13 & 3.12 & 1.02     \\\hline
\multirow{2}{*}{$\mathcal{P}$ = 0.5}  & 0.17 &  2.65 & 2.87 & -0.23 & 2.91 & 2.93 & -0.01 & 3.47 & 3.09 & 0.38 & 3.74 & 3.22 & 0.52     \\
              &     1.25  & 2.65 & 2.88 & -0.23 & 2.92 & 2.93 & -0.01 & 3.48 & 3.10 & 0.38 & 3.75 & 3.22 & 0.53     \\\hline
                     
 &  & \multicolumn{3}{c}{\texttt{OS1+UM1+SL1}}& \multicolumn{3}{c}{\texttt{OS1+UM10+SL1}} & \multicolumn{3}{c}{\texttt{OS1+UM50+SL1}} & \multicolumn{3}{c}{\texttt{OS1+UM100+SL1}} \\
\cmidrule{3-14}
& & ${\alpha}^{\,450\,\mu{\rm{m}}}_{\,850\,\mu{\rm{m}}}$ & ${\alpha}^{\,1000\,\mu{\rm{m}}}_{\,2000\,\mu{\rm{m}}}$ & $\Delta\,\alpha$ & ${\alpha}^{\,450\,\mu{\rm{m}}}_{\,850\,\mu{\rm{m}}}$ & ${\alpha}^{\,1000\,\mu{\rm{m}}}_{\,2000\,\mu{\rm{m}}}$ & $\Delta\,\alpha$ & ${\alpha}^{\,450\,\mu{\rm{m}}}_{\,850\,\mu{\rm{m}}}$ & ${\alpha}^{\,1000\,\mu{\rm{m}}}_{\,2000\,\mu{\rm{m}}}$ & $\Delta\,\alpha$ & ${\alpha}^{\,450\,\mu{\rm{m}}}_{\,850\,\mu{\rm{m}}}$ & ${\alpha}^{\,1000\,\mu{\rm{m}}}_{\,2000\,\mu{\rm{m}}}$ & $\Delta\,\alpha$  \\\hline
\multirow{2}{*}{Astrosil}   & 0.17 &2.07  & 2.75 & -0.68  & 2.15      & 2.22      & -0.07    & 2.17      & 2.07      & 0.1      & 2.18       & 2.05      & 0.13     \\
 & 1.25 &2.08  & 2.75 & -0.67   & 2.15      & 2.22      & -0.07    & 2.18      & 2.08      & 0.1      & 2.19       & 2.06      & 0.13     \\\hline
\multirow{2}{*}{$\mathcal{F_{\rm{ice}}}$ = 0.5}   & 0.17  & 2.55 & 2.82 & -0.26 & 2.26 & 2.29 & -0.03    & 2.17 & 2.19 & -0.02     & 2.15 & 2.17 & -0.02     \\
                     & 1.25  &2.56 & 2.82 & -0.26 & 2.26 & 2.29 & -0.03    & 2.17 & 2.19 & -0.02     & 2.16 & 2.17 & -0.01     \\\hline
\multirow{2}{*}{$\mathcal{P}$ = 0.5}  & 0.17  & 2.65  & 2.87 &  -0.23  & 2.24  & 2.33 &  -0.08   & 2.13  & 2.23 &  -0.10   & 2.11  & 2.22 &  -0.11     \\
                     & 1.25 & 2.65  & 2.88 &  -0.23   & 2.25  & 2.33 &  -0.08   & 2.13  & 2.23 &  -0.10   & 2.13  & 2.23 &  -0.10    \\\hline
\end{tabular}
\label{table:ice_porosity_SED with broken power law grain size distribution: over-under-abundance of small grains}
\end{table*}

\begin{figure*}[h!]
\centerline\centering\includegraphics[width=18cm, height=12cm]{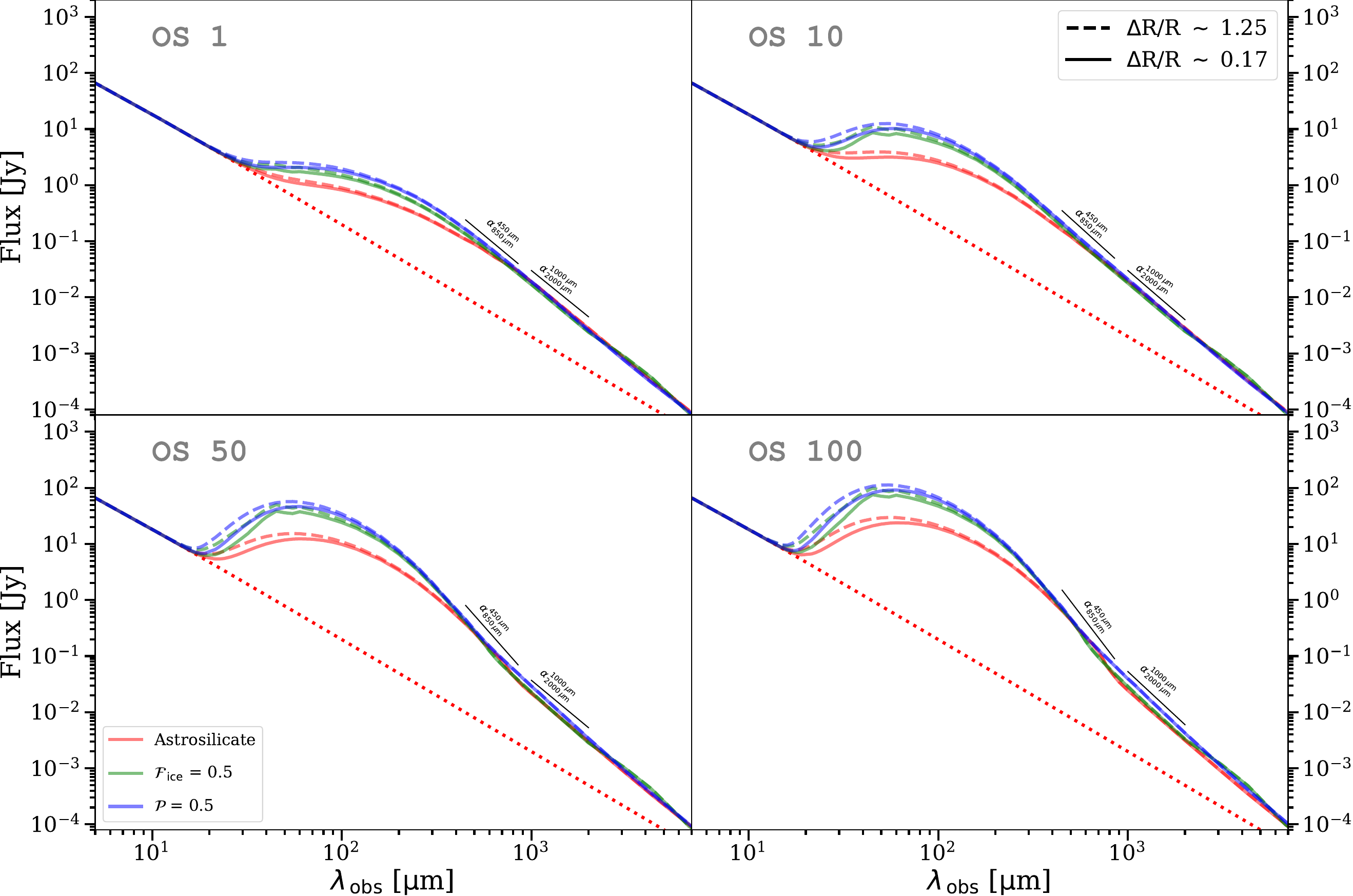}
\caption{SEDs of the case of the overabundance of small grains depending on the dust composition (ice inclusion $\mathcal{F}_{\rm ice}$ = 0.5 and porosity $\mathcal{P}$ = 0.5. Solid and dashed lines indicate the modeled SED of the dust reemission for the two different disk structures (thin belts and broad disks). The dotted lines represent the direct stellar radiation from A star. The top left plot shows the result based on the single power law grain size distribution (\texttt{OS1+UM1+SL1}), while the other three plots exhibit the results based on the broken power law grain size distribution, specifically the overabundance of small grains with the different scaling factors, 10, 50, and 100 (i.e., \texttt{OS10}, \texttt{OS50}, and \texttt{OS100}). See Sect.~\ref{res: Broken power law} and Table~\ref{table:ice_porosity_SED with broken power law grain size distribution: over-under-abundance of small grains} for details.}
\label{fig:SED_ice_porosity_OS}
\end{figure*}

\begin{figure*}
\centerline\centering\includegraphics[width=18cm, height=12cm]{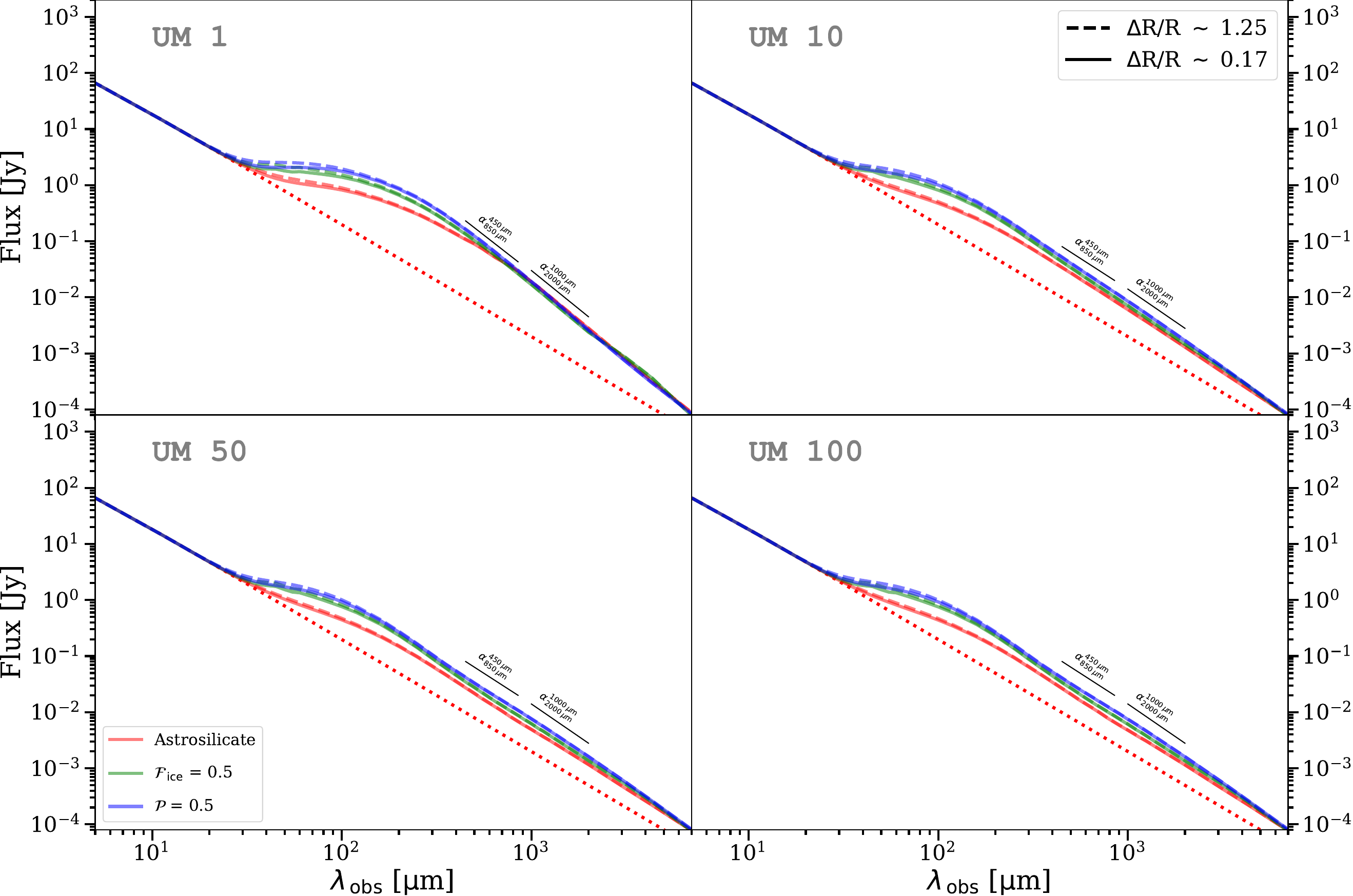}
\caption{SEDs of the case of the overabundance of small grains depending on the dust composition (ice inclusion $\mathcal{F}_{\rm ice}$ = 0.5 and porosity $\mathcal{P}$ = 0.5. Solid and dashed lines indicate the modeled SED of the dust reemission for the two different disk structures (thin belts and broad disks). The dotted lines represent the direct stellar radiation from A star. The top left plot shows the result based on the single power law grain size distribution (\texttt{OS1+UM1+SL1}), while the other three plots exhibit the results based on the broken power law grain size distribution, specifically the underabundance of medium grains with the different scaling factors, 10, 50, and 100 (i.e., \texttt{UM10}, \texttt{UM50}, and \texttt{UM100}). See Sect.~\ref{res: Broken power law} and Table~\ref{table:ice_porosity_SED with broken power law grain size distribution: over-under-abundance of small grains} for details.}
\label{fig:SED_ice_porosity_UM}
\end{figure*}

Variations in dust grain compositions, such as the inclusion of ice or consideration of porosity, affect the absorption and scattering efficiencies (\citealp{Kim2019}), thereby influencing the overall shape of SED (\citealp{Lebreton2012, Schueppler2015}). To understand the effect of dust compositions on the spectral inversion and potential degeneracy, we examine the implications of altering chemical compositions, specifically by integrating crystalline water ice into icy-astrosilicate mixtures (e.g., 50\% ice volume fraction ${\mathcal{F}}_{\rm ice}$~=~0.5) and the microstructure of the dust grains with introducing moderate porosity (e.g., 50\% porosity $\mathcal{P}$~=~0.5) through effective medium approximations (e.g., Maxwell-Garnett rule; \citealp{Maxwell-Garnett1904}) based on the grain size distribution described by the single power law and broken power law.

In cases of single power law grain size distribution and a lower over/underabundance of small/medium grains of the broken power law grain size distribution (e.g., \texttt{OS10+UM1+SL1} and \texttt{OS1+UM10+SL1}; see the first and second columns of Table~\ref{table:ice_porosity_SED with broken power law grain size distribution: over-under-abundance of small grains}, and Figs. \ref{fig:SED_ice_porosity_OS} and \ref{fig:SED_ice_porosity_UM}), resulting in a decrease in the difference in their slopes (i.e., smaller $\Delta\alpha$) and might help the situation. However, these modifications not only steepen the slope in the far-IR range but also affect the mm range, resulting in relatively small spectral inversions. Furthermore, both models lead to even smaller deviations in the spectral slopes, particularly with an increase in over/underabundance of small/medium grains based on the broken power law grain size distribution model (e.g., \texttt{OS50+UM1+SL1} and \texttt{OS1+UM50+SL1}; see the third and fourth columns of Table~\ref{table:ice_porosity_SED with broken power law grain size distribution: over-under-abundance of small grains}, and Figs. \ref{fig:SED_ice_porosity_OS} and \ref{fig:SED_ice_porosity_UM}). Consequently, these modifications to dust compositions alone are insufficient to reproduce the observed spectral inversion. This indicates that the specific composition is not the sole factor responsible for the spectral inversions seen in the far-infrared to mm ranges debris disk SEDs.

Overall, the effect of ice inclusion seems stronger than that of porosity in cases of an increase in the over/underabundance of small/medium grains. On the other hand, for the case of underabundance of medium grains, all of our models with the modification of dust composition fail to reproduce spectral inversion, highlighting the more significant effect of a higher abundance of small grains. Moreover, the fractional width does not seem to play a significant role in these outcomes, suggesting that the spectral characteristics are largely independent of this aspect of grain size distribution. This finding is consistent with our previous results in Sect.~\ref{Results}.

\subsection{Possible connection between the inferred size distribution and the collisional state of the disk}\label{Dis: Possible connection between the inferred size distribution and the collisional state of the disk}

As \citet{MacGregor2016} suggested, the slope of the grain size distribution $\gamma$ can be derived from the measurement of the mm spectral index $\alpha_{\rm mm}$ based on the dust opacity power law index of small particles (e.g., compared to observing wavelengths) $\beta_{\rm s}$ and an inferred power-law exponent from the dust temperature for each disk (i.e., spectral index of the Planck function $\alpha_{\rm PL}$) given wavelength ranges. Since most astronomical dust compositions have very similar values of $\beta_{\rm s}$ (e.g., 1.5 - 2) and the dust temperature only affects the Planck function weakly at long wavelengths (e.g., $\alpha_{\rm PL}$ $\sim$ 1.9 to 1.95 with the value of 2 as the Rayleigh-Jeans limit), $\alpha_{\rm mm}$ directly corresponds to the slope of the grain size distribution $\gamma$.

The inferred range of the parameters $\gamma$ based on $\alpha_{\rm mm}$ for the overabundance of small grains from our models (e.g., $\sim$ -3.5 to -3.6 regardless of stellar properties and dust compositions) is in good agreement with the predictions of steady-state collisional models, both from the classical model and modern models (e.g., \citealp{Dohnanyi1969, Marshall2017}), while the one from underabundance of medium grains showing a bit shallower slopes of grain size distributions (e.g., $\gamma$ $\sim$ -3.0), which can be explained by the size dependence of the forces removing the smaller dust grains from the system (e.g., PR drag and stellar radiation).

These observational results have yielded valuable insights, yet they also highlight the intrinsic challenges faced in such measurements, primarily due to the typical uncertainties (e.g., around 5 to 15\%) present in observational data, leading to spectral inversions that differ by approximately $\sim$ 10\%. These differences may stem from the varying sensitivities of telescopes, along with the diverse calibration and reduction techniques employed to mitigate atmospheric conditions, optical flaws, and environmental sensitivity, all of which impact the accuracy and precision of observations. Additionally, our studies show variability in spectral indices, which appears to depend on the specific properties of the considered debris disks (see Sect.~\ref{Results}). The complexity of these factors underscores advanced solutions, which are anticipated to be addressed by the ngVLA (Next Generation Very Large Array; \citealt{DiFrancesco2019}).

\subsection{Effect of the blow-out size dependent grain size distribution on the SEDs}\label{Effect of the blow-out size dependent grain size distribution on the SEDs}

\begin{figure*}[h!]
\centerline\centering\includegraphics[width=18cm, height=12cm]{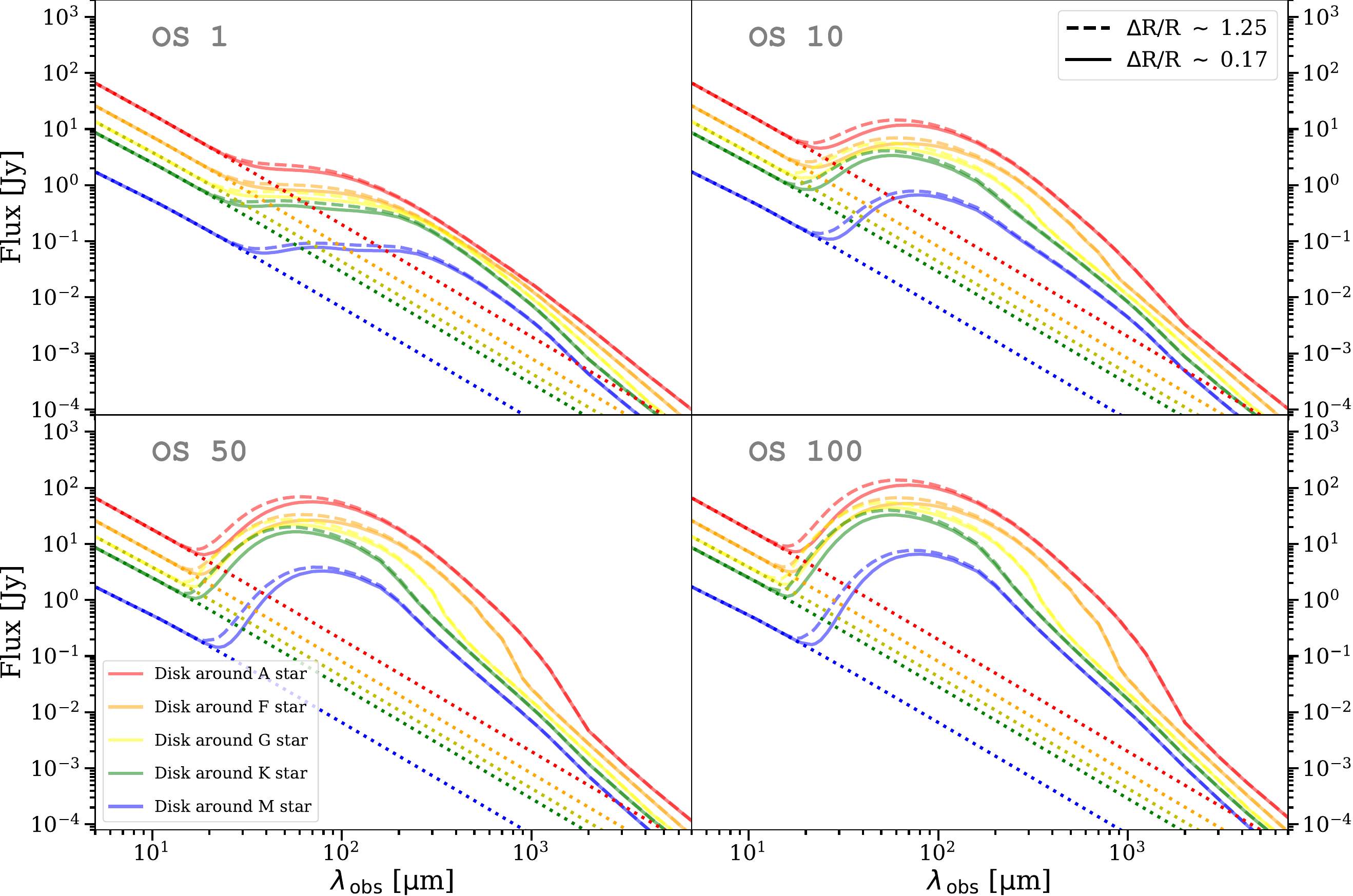}
\caption{SEDs of the case of the overabundance of small grains depending on the stellar properties (i.e., blow-out sizes). Solid and dashed lines indicate the modeled SED of the dust reemission for the two different disk structures (thin belts and broad disks). The dotted lines represent the direct stellar radiations. The top left plot shows the result based on the single power law grain size distribution (\texttt{OS1+UM1+SL1}), while the other three plots exhibit the results based on the broken power law grain size distribution, specifically the overabundance of small grains with the different scaling factors, 10, 50, and 100 (i.e., \texttt{OS10}, \texttt{OS50}, and \texttt{OS100}). See Table~\ref{GD_bo_star} and Fig.~\ref{fig:bo_GDs} for the proposed grain size distributions depending on the central star and blow-out sizes (\citealp{Thebault_Augereau_2007, Kim2018}).}
\label{fig:bo_appen_SED_over_small}
\end{figure*}

\begin{figure*}
\centerline\centering\includegraphics[width=18cm, height=12cm]{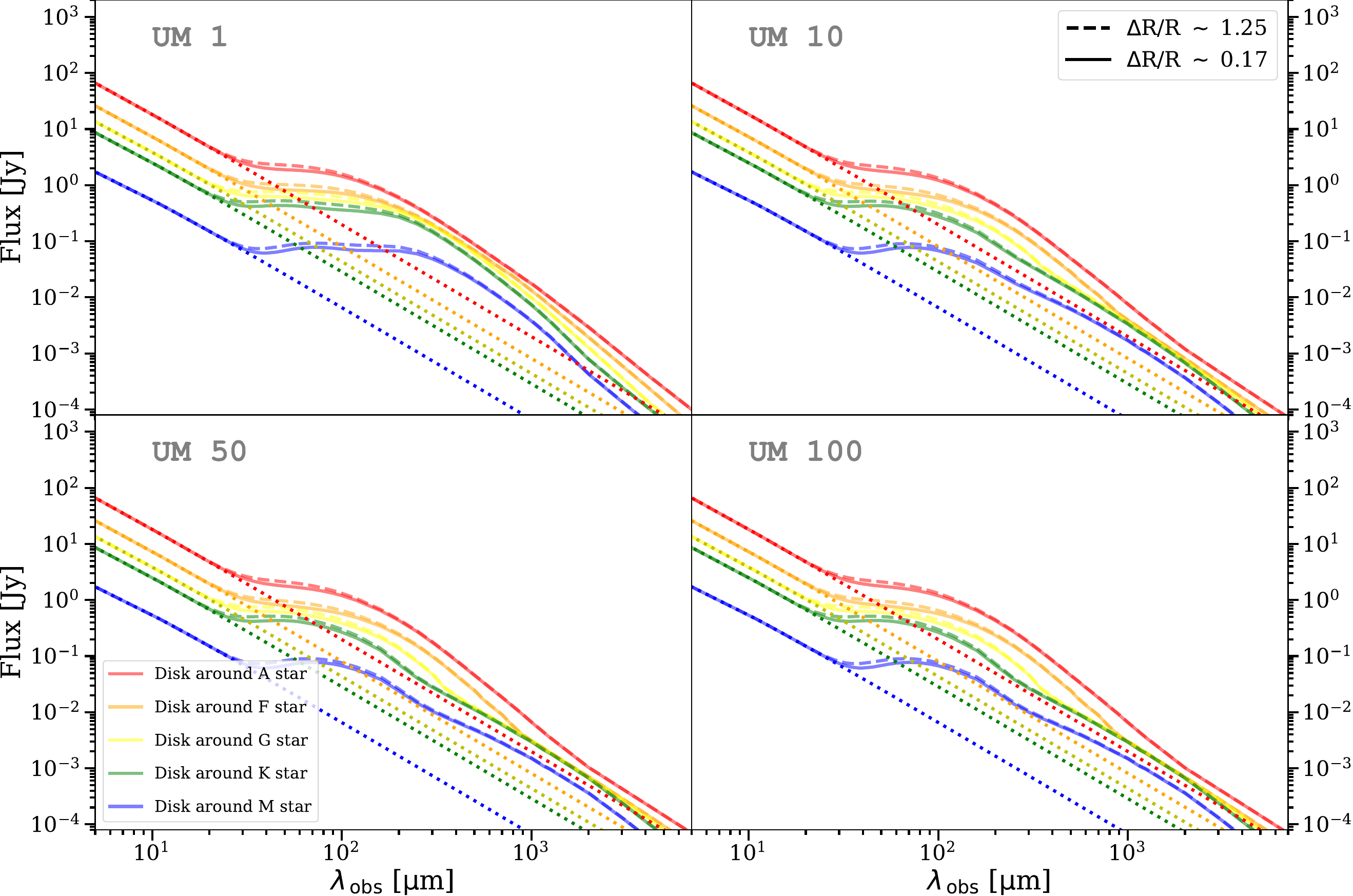}
\caption{SEDs of the case of the underabundance of medium grains depending on the stellar properties (i.e., blow-out sizes). Solid and dashed lines indicate the modeled SED of the dust reemission for the two different disk structures (thin belts and broad disks). The dotted lines represent the direct stellar radiations. The top left plot shows the result based on the single power law grain size distribution (\texttt{OS1+UM1+SL1}), while the other three plots exhibit the results based on the broken power law grain size distribution, specifically the underabundance of medium grains with the different scaling factors, 10, 50, and 100 (i.e., \texttt{UM10}, \texttt{UM50}, and \texttt{UM100}). See Table~\ref{GD_bo_star} and Fig.~\ref{fig:bo_GDs} for the proposed grain size distributions depending on the central star and blow-out sizes (\citealp{Thebault_Augereau_2007, Kim2018}).}
\label{fig:bo_SED_under_medium}
\end{figure*}

We also consider the grain size distribution with the overabundance and/or underabundance of different grain sizes as a function of stellar properties (i.e., blow-out size; see also Sect. \ref{Broken power law GD}). We find that the variations in the blowout size and spectral type that are responsible for the location of the first and second maximum/minimum in the grain size distribution (see also Fig. \ref{fig:bo_GDs} and Table \ref{GD_bo_star}), in turn, change the spectral slope of the SED at varying wavelength regimes (Figs.~\ref{fig:bo_appen_SED_over_small} and \ref{fig:bo_SED_under_medium}). For example, a grain size distribution with an overabundance of small grains in the case of an A-type star displays a change of the slope at significantly longer wavelengths (e.g., $\sim$ a few mm). On the other hand, K and M-type stars show this impact at even shorter wavelengths (e.g., $\sim$ a few hundred micrometers; see Fig.~\ref{fig:bo_appen_SED_over_small}). In the case of an F-type star, the results are comparable to those discussed in previous Sect.~\ref{Broken power law GD} because of similar grain size distribution. These results illustrate that the location of the gap in the grain size distribution has a direct impact on the wavelength of which the slope of SED is affected. This finding highlights the impact of the blow-out size on the resulting -even sub-mm/mm- slope of the SED. Furthermore, we find that the underabundance of medium-sized grains has a rather weak impact on the changing slope of the SED (see Figs.~\ref{fig:bo_appen_SED_over_small} and \ref{fig:bo_SED_under_medium}; Tables \ref{table:bo_SED with broken power law grain size distribution: overabundance of small grains} and
\ref{table:bo_SED with broken power law grain size distribution: underabundance of mid grains}), which is consistent with our previous findings discussed in Sect. \ref{res: Broken power law}.

The results discussed in this section suggest the need to constrain the slope of the SED in the different wavelength regimes to disentangle the impact of the spectral type from the impact of potentially, other physical mechanisms on the grain size distribution and thus the slope of the SED. However, we note that previous observational studies (e.g., \citealp{Lestrade2020, MacGregor2016}; see also Sect.~\ref{Dis: Application to observations}) have not identified a clear dependence of the impact of stellar spectral type on the slope of the SED, showing consistent observed spectral breaks occurring at similar wavelength regions, irrespective of the central star types.

\begin{table*}[]
\small
\centering
\def\arraystretch{1.5}   
\caption{Comparison of the spectral index of the proposed discontinuous grain size distributions (see Table~\ref{GD_bo_star} and Fig. \ref{fig:bo_GDs}) depending on the overabundance of small grains, stellar parameters (e.g., stellar temperature and radius), and disk parameters (e.g., fractional width $\Delta$$R$/$R$). For the locations of the first and second maximum/minimum in grain size distribution depending on the central star for the simulations see Table~\ref{GD_bo_star}.}
\begin{tabular}{llccc|cccccccccc}
\hline\hline
\multirow{3}{*}{SpT} & \multirow{3}{*}{$\Delta$$R$/$R$} & \multicolumn{3}{c}{Single power law} & \multicolumn{9}{c}{overabundance of small grains} \\\cmidrule{3-14}
 &  & \multicolumn{3}{c}{\texttt{OS1+UM1+SL1}}& \multicolumn{3}{c}{\texttt{OS10+UM1+SL1}} & \multicolumn{3}{c}{\texttt{OS50+UM1+SL1}}& \multicolumn{3}{c}{\texttt{OS100+UM1+SL1}}\\
\cmidrule{3-14}
&  & ${\alpha}^{\,450\,\mu{\rm{m}}}_{\,850\,\mu{\rm{m}}}$ & ${\alpha}^{\,1000\,\mu{\rm{m}}}_{\,2000\,\mu{\rm{m}}}$ & $\Delta\,\alpha$ &${\alpha}^{\,450\,\mu{\rm{m}}}_{\,850\,\mu{\rm{m}}}$ & ${\alpha}^{\,1000\,\mu{\rm{m}}}_{\,2000\,\mu{\rm{m}}}$ & $\Delta\,\alpha$ & ${\alpha}^{\,450\,\mu{\rm{m}}}_{\,850\,\mu{\rm{m}}}$ & ${\alpha}^{\,1000\,\mu{\rm{m}}}_{\,2000\,\mu{\rm{m}}}$ & $\Delta\,\alpha$ & ${\alpha}^{\,450\,\mu{\rm{m}}}_{\,850\,\mu{\rm{m}}}$ & ${\alpha}^{\,1000\,\mu{\rm{m}}}_{\,2000\,\mu{\rm{m}}}$ & $\Delta\,\alpha$  \\\hline
\multirow{2}{*}{A}   & 0.17    & 2.33  & 2.58 & -0.25  & 3.03  & 3.67 & -0.63  & 3.25  & 4.97 & -1.73  & 3.28  & 5.43 & -2.15 \\
& 1.25  & 2.34  & 2.58 & -0.25 & 3.04  & 3.67 & -0.63  & 3.26  & 4.98 & -1.72  & 3.29  & 5.44 & -2.15 \\\hline
\multirow{2}{*}{F}   & 0.17   & 2.27  & 2.76 & -0.50  & 3.61  & 2.89 & 0.72  & 4.60  & 3.24 & 1.36  & 4.84  & 3.45 & 1.39 \\
                     & 1.25 & 2.28  & 2.77 & -0.49  & 3.62  & 2.90 & 0.72  & 4.62  & 3.24 & 1.37  & 4.85  & 3.46 & 1.40     \\\hline
\multirow{2}{*}{G}   & 0.17   & 2.56  & 2.93 & -0.37  & 2.74  & 2.98 & -0.24  & 3.14  & 3.13 & 0.01  & 3.33  & 3.23 & 0.10\\
                     & 1.25  & 2.58  & 2.94 & -0.36  & 2.76  & 2.99 & -0.23  & 3.15  & 3.13 & 0.01  & 3.34  & 3.23 & 0.11 \\\hline
\multirow{2}{*}{K}   & 0.17  &  2.59  & 3.21 & -0.62  & 2.73  & 3.24 & -0.51  & 3.03  & 3.32 & -0.29  & 3.18  & 3.36 & -0.19 \\
                     & 1.25  & 2.61  & 3.22 & -0.61  & 2.74  & 3.25 & -0.51  & 3.04  & 3.32 & -0.29  & 3.18  & 3.37 & -0.19     \\\hline
\multirow{2}{*}{M}   & 0.17  &  2.29  & 3.14 & -0.85  & 2.50  & 3.18 & -0.67  & 2.89  & 3.27 & -0.38  & 3.06  & 3.32 & -0.26\\
                     & 1.25  & 2.33  & 3.15 & -0.82  & 2.53  & 3.18 & -0.66  & 2.90  & 3.27 & -0.37  & 3.06  & 3.32 & -0.26
    \\\hline
\end{tabular}
\label{table:bo_SED with broken power law grain size distribution: overabundance of small grains}
\end{table*}

\begin{table*}[]
\centering
\small
\def\arraystretch{1.5}   
\caption{Comparison of the spectral index of the proposed discontinuous grain size distributions (see Table~\ref{GD_bo_star} and Fig. \ref{fig:bo_GDs}) depending on the underabundance of medium grains, stellar parameters (e.g., stellar temperature and radius), and disk parameters (e.g., fractional width $\Delta$$R$/$R$). For the locations of the first and second maximum/minimum in grain size distribution depending on the central star for the simulations see Table~\ref{GD_bo_star}.}
\begin{tabular}{llccc|cccccccccc}
\hline\hline
\multirow{3}{*}{SpT} & \multirow{3}{*}{$\Delta$$R$/$R$} & \multicolumn{3}{c}{Single power law} & \multicolumn{9}{c}{underabundance of medium grains} \\\cmidrule{3-14}
 &  & \multicolumn{3}{c}{\texttt{OS1+UM1+SL1}}& \multicolumn{3}{c}{\texttt{OS1+UM10+SL1}} & \multicolumn{3}{c}{\texttt{OS1+UM50+SL1}} & \multicolumn{3}{c}{\texttt{OS1+UM100+SL1}} \\
\cmidrule{3-14}
& & ${\alpha}^{\,450\,\mu{\rm{m}}}_{\,850\,\mu{\rm{m}}}$ & ${\alpha}^{\,1000\,\mu{\rm{m}}}_{\,2000\,\mu{\rm{m}}}$ & $\Delta\,\alpha$ & ${\alpha}^{\,450\,\mu{\rm{m}}}_{\,850\,\mu{\rm{m}}}$ & ${\alpha}^{\,1000\,\mu{\rm{m}}}_{\,2000\,\mu{\rm{m}}}$ & $\Delta\,\alpha$ & ${\alpha}^{\,450\,\mu{\rm{m}}}_{\,850\,\mu{\rm{m}}}$ & ${\alpha}^{\,1000\,\mu{\rm{m}}}_{\,2000\,\mu{\rm{m}}}$ & $\Delta\,\alpha$ & ${\alpha}^{\,450\,\mu{\rm{m}}}_{\,850\,\mu{\rm{m}}}$ & ${\alpha}^{\,1000\,\mu{\rm{m}}}_{\,2000\,\mu{\rm{m}}}$ & $\Delta\,\alpha$  \\\hline
\multirow{2}{*}{A}   & 0.17  & 2.33  & 2.58 &  -0.25  & 2.68  & 2.66 &  0.02  & 2.74  & 2.68 &  0.07  & 2.75  & 2.68 &  0.07\\
 & 1.25  & 2.34  & 2.58 &  -0.25  & 2.69  & 2.66 &  0.02  & 2.75  & 2.68 &  0.07  & 2.76  & 2.68 &  0.08     \\\hline
\multirow{2}{*}{F}   & 0.17 & 2.27  & 2.76 &  -0.50  & 2.87  & 2.28 &  0.59  & 3.05  & 2.13 &  0.92  & 3.08  & 2.11 &  0.97\\ 
& 1.25  & 2.28  & 2.77 &  -0.49  & 2.88  & 2.28 &  0.60  & 3.07  & 2.13 &  0.94  & 3.10  & 2.11 &  0.99\\\hline
\multirow{2}{*}{G}   & 0.17   & 2.56  & 2.93 &  -0.37  & 2.13  & 2.31 &  -0.17  & 1.97  & 2.17 &  -0.20  & 1.94  & 2.15 &  -0.21\\ 
& 1.25   & 2.58  & 2.94 &  -0.36  & 2.14  & 2.31 &  -0.17  & 1.97  & 2.17 &  -0.20  & 1.95  & 2.15 &  -0.21    \\\hline
\multirow{2}{*}{K}   & 0.17   & 2.59  & 3.21 &  -0.62  & 2.06  & 2.29 &  -0.24  & 1.89  & 2.15 &  -0.27  & 1.86  & 2.13 &  -0.27     \\ & 1.25  & 2.61  & 3.22 &  -0.61  & 2.07  & 2.30 &  -0.23  & 1.90  & 2.16 &  -0.26  & 1.87  & 2.14 &  -0.27\\\hline
\multirow{2}{*}{M}   & 0.17  & 2.29  & 3.14 &  -0.85  & 1.80  & 2.20 &  -0.40  & 1.64  & 2.05 &  -0.41  & 1.62  & 2.04 &  -0.42\\  
& 1.25  & 2.33  & 3.15 &  -0.82  & 1.83  & 2.21 &  -0.38  & 1.67  & 2.06 &  -0.40  & 1.64  & 2.04 &  -0.40 \\ \hline
\end{tabular}
\label{table:bo_SED with broken power law grain size distribution: underabundance of mid grains}
\end{table*}

\section{Summary}\label{Summary and Conclusions}

Motivated by the expected deviation of the grain size distribution from the often assumed continuous distribution described by a single power law, we investigated the impact of the over- and underabundance of grains in specific size regimes on the SED of debris disks. Based on an exemplary, simple model, we quantified and discussed the impact of different values of over/underabundances of submicrometer to mm-sized grains in a debris dust ring with two different fractional widths $\Delta\,R/R$ around central stars with spectral types A, F, G, K, and M on the slope of the FIR to mm SED of these systems. We described this shape and its the spectral indices ${\alpha}^{450\,\mu{\rm{m}}}_{850\,\mu{\rm{m}}}$ and ${\alpha}^{1000\,\mu{\rm{m}}}_{2000\,\mu{\rm{m}}}$, and discussed their differences $\Delta\,\alpha$. Our key findings are:

\begin{itemize}

\item [--] The deviation of a grain size distribution from that described by a single power law can be potentially derived from the analysis of the SED in the FIR to mm wavelength range. More specifically, the characteristics of the slope of the SED via spectral indices and wavelengths at which photometric data can be obtained even from ground-based telescopes appear promising. In the considered model, characterized by a chosen grain size distribution, i.e., the specific size range of grains for which an over/underabundance was assumed, we find a characteristic kink in the SED around 800 to 900 $\mu$m. \newline

\item [--] As the change of the slope and the wavelength range of its occurrence is directly linked to the deviation of the grain size distribution from a single power law distribution, these observations potentially provide the key for constraining the underlying physical processes leading to the specific grain size distribution. In particular, these observations provide the potential to constrain the collisional physics (e.g., material strength) of the otherwise hardly accessible properties of the larger grains ($\gg$ cm). \newline

\item [--] The presence of the overabundance of small grains and the underabundance of medium grains both contribute to a steeper slope in the SED at the far-IR range. However, their effects differ at mm wavelengths. Specifically, the overabundance of small grains primarily results in a much higher value ${\alpha}^{\,450\,\mu{\rm{m}}}_{\,850\,\mu{\rm{m}}}$ (i.e., steeper slope), while the ${\alpha}^{\,1000\,\mu{\rm{m}}}_{\,2000\,\mu{\rm{m}}}$ becomes slightly higher (i.e, slightly steeper slope). On the other hand, the underabundance of medium grains leads to a higher value of ${\alpha}^{\,450\,\mu{\rm{m}}}_{\,850\,\mu{\rm{m}}}$ (i.e, slightly steeper slope), but primarily results in a lower value of ${\alpha}^{\,1000\,\mu{\rm{m}}}_{\,2000\,\mu{\rm{m}}}$ (i.e., shallower slope). The impact of the overabundance of small dust particles is stronger than that of the underabundance of medium-sized dust particles. \newline

\item [--] We found that the spectral index transition is more pronounced for debris disks around brighter, i.e., early-type central stars and broader disks. However, the disk structure such as the fractional width of the disk system leaves only little effect on the FIR to mm slope of the SED if compared to the impact of the grain size distributions. \newline

\item [--] We found that dust composition is not the sole physical mechanism responsible for the spectral inversion observed in the far-IR to (sub-)mm part of the SEDs of debris disk systems.\newline

\item [--] If we assume that the location of the first and second maximum/minimum of the grain size distribution depends on the blow-out size, and thus on the stellar spectral type, the corresponding impact on the sub-mm/mm on SED is wavelength-dependent. \newline

\end{itemize}


\begin{acknowledgements}
The authors thank J.-F. Lestrade for fruitful discussions on the modeling of specific grain size distribution in this study. MK gratefully acknowledges the funding from the Royal Society. MK and SW gratefully acknowledge the financial support by the Deutsche Forschungsgemeinschaft under contracts \emph{WO 857/15-1} and \emph{WO 857/15-2} (DFG). 
\end{acknowledgements}

\bibliographystyle{aa}
\bibliography{bib_impact_of_grain_on_debris_disks.bib}

\twocolumn
\begin{appendix}

\end{appendix}

\end{document}